# Generalized Averaging Method for Power Electronics Modeling from DC to above Half the Switching Frequency

Hongchang Li, *Senior Member, IEEE*, Kangping Wang, *Senior Member, IEEE*, Jingyang Fang, *Senior Member, IEEE*, Wenjie Chen, *Senior Member, IEEE*, and Xu Yang, *Senior Member, IEEE*

*Abstract*- Modeling power electronic converters at frequencies close to or above half the switching frequency has been difficult due to the time-variant and discontinuous switching actions. This paper uses the properties of moving Fourier coefficients to develop the generalized averaging method, breaking though the limit of half the switching frequency. The paper also proposes the generalized average model for various switching signals, including pulse-width modulation (PWM), phase-shift modulation, pulse-frequency modulation (PFM), and state-dependent switching signals, so that circuits and modulators/controllers can be modeled separately and combined flexibly. Using the Laplace transform of moving Fourier coefficients, the coupling of signals and their sidebands at different frequencies is clearly described as the coupling of moving Fourier coefficients at the same frequency in a linear time-invariant system framework. The modeling method is applied to a PWM controlled boost converter, a $V^2$ constant on-time controlled buck converter, and a PFM controlled LLC converter, for demonstration and validation. Experimental results of the converters in different operating modes show that the proposed models have higher accuracy than exiting models, especially in the frequency range close to or above half the switching frequency. The developed method can be applied to almost all types of power electronic converters.

## I. Introduction

Power electronic converters play an increasingly important role in today's energy, transportation, and information industries. Modeling and control of these converters are critical for the stable and efficient operation [1-4]. The best known and most commonly used method for modeling power electronic converters is the state-space averaging [5]. The method overcomes the time-variations and discontinuities caused by switching actions by averaging the exact state-space descriptions of switched models over a single switching cycle, and a linear time-invariant (LTI) small-signal version can be further derived through local linearization [6]. However, the averaged model is effective only when the frequency of interest is well below half the switching frequency. Characterizing the dynamics at frequencies close to or above half the switching frequency is also important for fast control and stability analysis, but it has been difficult.

To obtain an accurate picture of closed-loop behavior near half the switching frequency, Packard [7] presented a discrete modeling method for switching regulators. The method integrates the small-signal perturbation equations across a complete switching cycle including modulation to find the small-signal difference equations. The integration is easy in principle but difficult to implement because the computation of the exponential matrices in the difference equations is a formidable task. To overcome this difficulty, Packard introduced the "straight-line" approximation, which states that, the exponential matrices describing the evolution of states can be accurately represented, in their intervals of validity, by the first two terms of their Taylor series expansions, with all higher order terms neglected. Brown and Middlebrook [8] found that the "straight-line" approximation of the difference equations is precisely the "straight-line" approximation of the averaged differential equations, and proposed the sampled-data modeling method, which incorporates both the continuous form of the averaged model and the high-frequency accuracy of the discrete model. The most representative application of the sampled-data model is the current mode control [9-11], where subharmonic instabilities can be predicted by the model. Although the discrete model and the sampled-data model are accurate at high frequencies, the modeling relies on the calculation of switching instants, which is often very complicated, especially when the switching instants are implicitly controlled by state trajectories [12]. Besides, the mixing of difference equations or sampler models with the continuous model of the rest of the converter system, e.g., the analog feedback and compensator, also causes inconvenience in practical applications.

From the perspective of control theory, the steady state of a power electronic converter is a limit cycle on its state space, and can be analyzed by the describing function method, which was developed in the 1930s to solve certain nonlinear control problems [13]. The calculation of a describing function is based on the harmonic balance principle and only uses the small-signal assumption. Therefore, the describing function method is valid at any high frequency, and is also applicable to the current-mode control [14, 15]. However, due to the periodic switching, describing functions of a power electronic converter can only be determined over a long commensurate period [16], which is an integer multiple of both the switching period and the modulation period, through a lengthy derivation of the Jacobian matrix of Fourier coefficients of waveforms [17]. During the derivation, many assumptions need to be made to simplify the geometry of the waveforms to find the switching boundaries between numerous time intervals [15]. These assumptions may introduce errors and limit the applicability of the method.

To simplify the implementation of the describing function method, Yang [16] proposed the extended describing function (EDF) method for the modeling of resonant converters. The method approximates the Jacobian matrix of Fourier coefficients defined over the long commensurate period by its counterpart defined over a short steady-state switching period. The approximation is made under two conditions: 1) the perturbation is at small-signal level, and 2) the modulation frequency is much lower than the switching frequency. The EDF method has become popular in the modeling of resonant converters due to its simplicity. However, its valid range was limited to less than half the switching frequency. Also for the modeling of resonant converters, Sanders et al. [18] proposed the generalized averaging method based on moving Fourier coefficients. Although the model is derived directly from the exact state-space model, its small-signal form is equivalent to the EDF model, therefore, the generalized averaging method is also considered valid only below half the switching frequency. In theory, both the EDF method and the generalized averaging method can improve their accuracy by taking into account more harmonic components, however, almost all applications only use DC and the fundamental harmonic, and in these cases, the methods are equivalent to the phasor transformation [19].

A model that successfully uses multiple harmonic components is the multifrequency model of buck converters [20, 21] presented by Qiu et al. The model considers the sidebands generated by the pulse-width-modulation (PWM) comparator and describes the coupling of the sidebands in a closed voltage loop. The model is effective above half the switching frequency and gives an intuitive frequency-domain explanation of the sideband effect. One application of the model is the study of current sharing of point-of-load converters under rapid load changes [22]. As an extension of the multifrequency model, Yue et al. [23] applied the harmonic state-space (HSS) model to power electronic converters. The HSS model was proposed by Wereley [24] based on the linear time-periodic (LTP) system theory. It assumes that the inputs, states, and outputs of an LTP system are all exponentially modulated periodic (EMP) signals, and finds their relationships using the harmonic balance principle. The EMP assumption implies that the HSS model is essentially a small-signal frequency-domain model, as is the multifrequency model, although they may exhibit a large-signal form for buck converters (because buck converters have a special large-signal linear topology). In a canonical HSS model, the periodic state matrix, input matrix, output matrix, and feedthrough matrix are independent of inputs and states, but this is not the case in power electronic converters since these matrices are functions of switching signals, and the switching signals usually depend on inputs and states. To be applied to power electronic converters, the HSS model must include a closed-loop control model that uses inputs and states to represent the switching signals and separate them from the matrices through local linearization [25-27]. Similarly, Qiu's model also needs to include a feedback transfer function from the output voltage to the modulator input. The inclusion of closed-loop control complicates the modeling process and makes the method impractical for more complex topologies with state-dependent switching. So far, the multifrequency model and the HSS model have only been applied to continuous-conduction-mode (CCM) buck-type converters, such as modular multilevel converters [27] and voltage-sourced inverters [28, 29].

Although many modeling methods have been proposed for power electronic converters and have been successful in their respective applications, the modeling process often requires piecewise waveform analysis and specific approximation techniques, leading to complexities and uncertainties. Besides, switching signals under various modulation schemes still lack a consistent and accurate description. Modeling the high-frequency characteristics under operating conditions such as the discontinuous conduction mode (DCM), variable switching frequency, and multiple resonances, remains challenging.

The purpose of this study is to find a more accurate modeling method suitable for different types of converters, and to establish a more solid foundation for the time-frequency analysis of power electronic systems. The work can be seen as a continuation of the generalized averaging method, and the contributions include:

1) The generalized averaging method is developed to include infinite harmonics in a concise formula, so that the effective range of the model breaks though the limit of half the switching frequency.
2) The generalized average model of switching signals is proposed, allowing various modulation/control schemes to be accurately modeled and flexibly combined with different circuit topologies.
3) The coupling of signals and their sidebands at different frequencies is clearly and conveniently described as the coupling of moving Fourier coefficients at the same frequency in an LTI system framework.
4) Three representative converters are modeled as examples and the most accurate models to date are obtained.

Table 1 compares the differences between the developed generalized averaging method and other modeling methods.

Table 1. Comparison of modeling methods

| Method | Effective range | Reported Applications | Limitations |
|---|---|---|---|
| State-space averaging | Below half the switching frequency | PWM converters [5] | Limited effective range and applications |
| Discrete modeling / sampled-data modeling | From DC to above half the switching frequency | PWM and resonant converters [8, 12] | Complex calculation of the piecewise state trajectories; Errors due to geometric simplifications of waveforms |
| Describing function | From DC to above half the switching frequency | PWM converters [14, 15] | Lengthy derivation over the commensurate period; Errors due to geometric simplifications of waveforms. |
| EDF / generalized averaging (DC and fundamental harmonics) | Below half the switching frequency | PWM and resonant converters [16, 18, 19] | Limited effective range |
| Multi-frequency / HSS | From DC to above half the switching frequency | Buck-type converters [20, 23] | Limited applications; Truncation errors |
| Developed generalized averaging (this work) | From DC to above half the switching frequency | PWM and resonant converters (this paper) | Truncation errors (can tend to 0) |

The remainder of the paper is organized as follows. In Section II, moving Fourier coefficients and their properties are reviewed and introduced as a basis for the modeling. Section III proceeds to the modeling steps of power electronic circuits. Section IV presents the generalized average model of switching signals. Section V, VI, and VII build the models of the PWM controlled boost converter, the $V^2$ constant on-time controlled buck converter, and the PFM controlled LLC converter, respectively, for demonstration and validation. Conclusions are presented in Section VIII.

## II. Moving Fourier Coefficients and Their Properties

### A. Moving Fourier coefficients

In this paper, for any $n \in \mathbb{Z}$, the $n$th moving Fourier coefficient of a signal $x(t)$ is given by

$$\langle x \rangle_n(\tau) = \frac{1}{T} \int_{\tau - \frac{T}{2}}^{\tau + \frac{T}{2}} x(t) e^{-jn\omega t} dt \quad (1)$$

where $T$ is the fundamental period and $\omega$ is the fundamental frequency. In particular, $\langle x \rangle_0(\tau)$ is the moving average of $x(t)$.

If $x(t)$ is of bounded variation on $[\tau - T/2, \tau + T/2]$, then the time-variant Fourier series of $x(t)$ converges to the arithmetic mean of the left and right limits of $x(t)$ at each $t \in (\tau - T/2, \tau + T/2)$:

$$\sum_{n=-\infty}^{+\infty} \langle x \rangle_n(\tau) e^{jn\omega t} = \lim_{\varepsilon \to 0} \frac{x(t+\varepsilon) + x(t-\varepsilon)}{2} \quad (2)$$

In particular, if $x(t)$ is continuous at $t$, then

$$\sum_{n=-\infty}^{+\infty} \langle x \rangle_n(\tau) e^{jn\omega t} = x(t) \quad (3)$$

The example in Fig. 1 shows that a non-periodic signal can be accurately represented by its time-variant Fourier series. The moving Fourier coefficients are continuous everywhere and differentiable almost everywhere although the original signal is discontinuous.

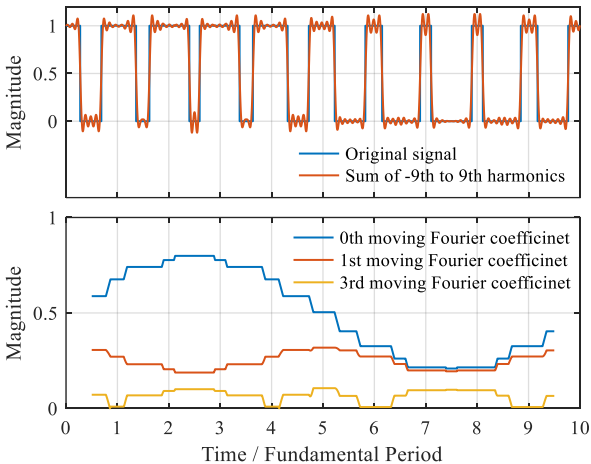

Fig. 1. A signal and some of its moving Fourier coefficients.

### B. Linearity

The moving Fourier coefficients of a linear combination of signals equal the same combination of their moving Fourier coefficients:

$$\langle ax + by \rangle_n(\tau) = \frac{1}{T} \int_{\tau-\frac{T}{2}}^{\tau+\frac{T}{2}} [ax(t) + by(t)] e^{-jn\omega t} dt$$
$$= a\langle x \rangle_n(\tau) + b\langle y \rangle_n(\tau) \quad (4)$$

where $a$ and $b$ are constants. In vector form:

$$\langle ax + by \rangle(\tau) = a\langle x \rangle(\tau) + b\langle y \rangle(\tau) \quad (5)$$

where "$\langle \cdot \rangle$" without a subscript represents the vector of sequentially arranged moving Fourier coefficients, e.g.:

$$\langle x \rangle(\tau) = \begin{bmatrix} \vdots \\ \langle x \rangle_{-1}(\tau) \\ \langle x \rangle_0(\tau) \\ \langle x \rangle_1(\tau) \\ \vdots \end{bmatrix} \quad (6)$$

### C. Multiplication

The moving Fourier coefficients of the multiplication of two signals equal the discrete convolution of their moving Fourier coefficients:

$$\langle xy \rangle_n(\tau) = \sum_{m=-\infty}^{+\infty} \langle x \rangle_{n-m}(\tau) \langle y \rangle_m(\tau) \quad (7)$$

In matrix form:

$$\langle xy \rangle(\tau) = \langle x \rangle(\tau) * \langle y \rangle(\tau) = [\![x]\!](\tau) \langle y \rangle(\tau) \quad (8)$$

where "$*$" is the convolution operator and "$[\![\cdot]\!]$" represents the Toeplitz matrix of the moving Fourier coefficients, e.g.:

$$[\![x]\!](\tau) = \begin{bmatrix} \ddots & \ddots & \ddots & \ddots & \ddots \\ \ddots & \langle x \rangle_0(\tau) & \langle x \rangle_{-1}(\tau) & \langle x \rangle_{-2}(\tau) & \ddots \\ \ddots & \langle x \rangle_1(\tau) & \langle x \rangle_0(\tau) & \langle x \rangle_{-1}(\tau) & \ddots \\ \ddots & \langle x \rangle_2(\tau) & \langle x \rangle_1(\tau) & \langle x \rangle_0(\tau) & \ddots \\ \ddots & \ddots & \ddots & \ddots & \ddots \end{bmatrix} \quad (9)$$

In particular, the Toeplitz matrices of the moving Fourier coefficients of constants 0 and 1 are $[\![0]\!] = O$ and $[\![1]\!] = I$, respectively, where $O$ is a zero matrix and $I$ is an identity matrix.

### D. Partial derivative

If $x(t)$, $f(x,t)$, and $\partial f / \partial x$, are all of bounded variation on $[\tau - T/2, \tau + T/2]$, then it is derived from the property of linearity and the property of multiplication that

$$d\langle f \rangle_n(\tau) = \langle df \rangle_n(\tau) = \langle \frac{\partial f}{\partial x} dx \rangle_n(\tau)$$
$$= \sum_{m=-\infty}^{+\infty} \langle \frac{\partial f}{\partial x} \rangle_{n-m}(\tau) \langle dx \rangle_m(\tau)$$
$$= \sum_{m=-\infty}^{+\infty} \langle \frac{\partial f}{\partial x} \rangle_{n-m}(\tau) d\langle x \rangle_m(\tau) \quad (10)$$

Consequently, for any pair of $n \in \mathbb{Z}$ and $m \in \mathbb{Z}$,

$$\frac{\partial \langle f \rangle_n}{\partial \langle x \rangle_m}(\tau) = \langle \frac{\partial f}{\partial x} \rangle_{n-m}(\tau) \quad (11)$$

In matrix form:

$$\frac{\partial\langle f\rangle}{\partial\langle x\rangle}(\tau) = \begin{bmatrix} \ddots & \ddots & \ddots & \ddots & \\ \ddots & \frac{\partial\langle f\rangle_{-1}}{\partial\langle x\rangle_{-1}}(\tau) & \frac{\partial\langle f\rangle_{-1}}{\partial\langle x\rangle_0}(\tau) & \frac{\partial\langle f\rangle_{-1}}{\partial\langle x\rangle_1}(\tau) & \ddots \\ \ddots & \frac{\partial\langle f\rangle_0}{\partial\langle x\rangle_{-1}}(\tau) & \frac{\partial\langle f\rangle_0}{\partial\langle x\rangle_0}(\tau) & \frac{\partial\langle f\rangle_0}{\partial\langle x\rangle_1}(\tau) & \ddots \\ \ddots & \frac{\partial\langle f\rangle_1}{\partial\langle x\rangle_{-1}}(\tau) & \frac{\partial\langle f\rangle_1}{\partial\langle x\rangle_0}(\tau) & \frac{\partial\langle f\rangle_1}{\partial\langle x\rangle_1}(\tau) & \ddots \\ & \ddots & \ddots & \ddots & \ddots \end{bmatrix}$$

$$= \left[\!\!\left[\frac{\partial f}{\partial x}\right]\!\!\right](\tau) \tag{12}$$

### E. Time derivative

If $x(t)$ is of bounded variation on $[\tau - T/2, \tau + T/2]$ and is continuous at $\tau - T/2$ and $\tau + T/2$, then $\langle x\rangle_n(\tau)$ is differentiable at $\tau$, and

$$\frac{d\langle x\rangle_n}{d\tau}(\tau) = \frac{d}{d\tau}\frac{1}{T}\int_{\tau-\frac{T}{2}}^{\tau+\frac{T}{2}} x(t)e^{-jn\omega t}dt$$

$$= \frac{1}{T}x(t)e^{-jn\omega t}\Big|_{\tau-\frac{T}{2}}^{\tau+\frac{T}{2}} \tag{13}$$

In addition, if $x'(t) = dx/dt$ is of bounded variation on $[\tau - T/2, \tau + T/2]$, then

$$\frac{d\langle x\rangle_n}{d\tau}(\tau) = \frac{1}{T}\int_{\tau-\frac{T}{2}}^{\tau+\frac{T}{2}} \frac{dx(t)e^{-jn\omega t}}{dt}dt$$

$$= \langle x'\rangle_n(\tau) - jn\omega\langle x\rangle_n(\tau) \tag{14}$$

In vector form:

$$\frac{d\langle x\rangle}{d\tau}(\tau) = \langle x'\rangle(\tau) - jN\omega\langle x\rangle(\tau) \tag{15}$$

where $N$ is a diagonal matrix that collects the indices of the moving Fourier coefficients:

$$N = \text{diag}(\cdots, -2, -1, 0, 1, 2, \cdots) \tag{16}$$

Furthermore, for any $t \in (\tau - T/2, \tau + T/2)$,

$$\sum_{n=-\infty}^{+\infty} jn\omega\langle x\rangle_n(\tau)e^{jn\omega t} = \lim_{\varepsilon\to 0}\frac{x'(t+\varepsilon) + x'(t-\varepsilon)}{2} \tag{17}$$

In particular, if $x'(t)$ is continuous at $t$, then

$$\sum_{n=-\infty}^{+\infty} jn\omega\langle x\rangle_n(\tau)e^{jn\omega t} = x'(t) \tag{18}$$

### F. Laplace transform

The Laplace transform of a moving Fourier coefficient equals the shifted and scaled Laplace transform of the original signal:

$$\mathcal{L}\{\langle x\rangle_n(\tau)\}(s) = \mathcal{L}\left\{\frac{1}{T}\int_{\tau-\frac{T}{2}}^{\tau+\frac{T}{2}} x(t)e^{-jn\omega t}dt\right\}(s)$$

$$= \frac{e^{\frac{T}{2}s} - e^{-\frac{T}{2}s}}{Ts}\mathcal{L}\{x(t)\}(s + jn\omega)$$

$$= \text{sinc}\frac{s}{j\omega}\mathcal{L}\{x(t)\}(s + jn\omega) \tag{19}$$

where the scaling factor is shown in Fig. 2.

If there is transfer function matrix $H(s)$ from $\langle x\rangle(\tau)$ to $\langle y\rangle(\tau)$, as shown in Fig. 3 (a), then for any $m \in \mathbb{Z}$,

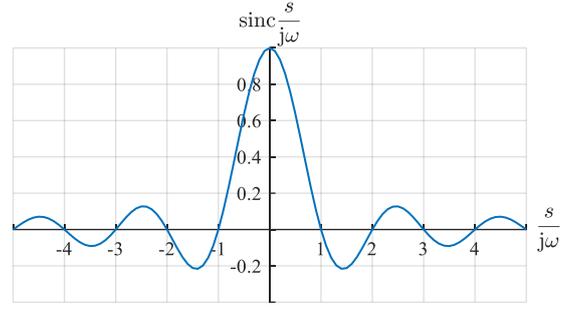

Fig. 2. The scaling factor.

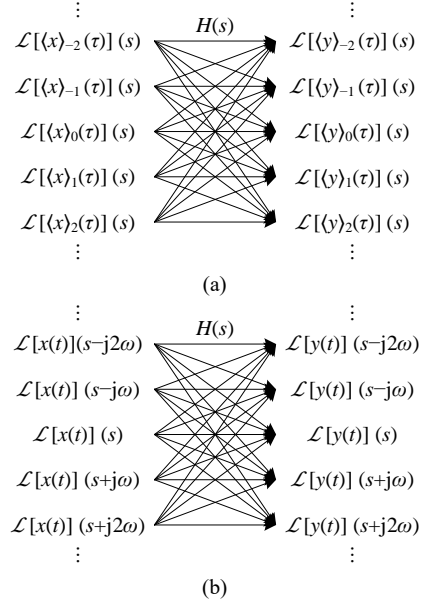

Fig. 3. Transfer function matrix between (a) moving Fourier coefficients and (b) signals and sidebands.

$$\mathcal{L}\{\langle y\rangle_m(\tau)\}(s) = \sum_{n=-\infty}^{+\infty} H_{mn}(s)\mathcal{L}\{\langle x\rangle_n(\tau)\}(s) \tag{20}$$

where $H_{mn}(s)$ is the $(m,n)$th element of $H(s)$. According to (19), for any $s$ that is not a non-zero integer multiple of $j\omega$, $H(s)$ is also the "transfer function" matrix from $x(t)$ and its sidebands to $y(t)$ and its sidebands, as shown in Fig. 3 (b), and therefore

$$\mathcal{L}\{y(t)\}(s + jm\omega) = \sum_{n=-\infty}^{+\infty} H_{mn}(s)\mathcal{L}\{x(t)\}(s + jn\omega) \tag{21}$$

If $x(t)$ does not contain the frequency component at $s + jn\omega$ for any non-zero integer $n$, then

$$\frac{\mathcal{L}\{y(t)\}(s + jm\omega)}{\mathcal{L}\{x(t)\}(s)} = H_{m0}(s) \tag{22}$$

In particular

$$\frac{\mathcal{L}\{y(t)\}(s)}{\mathcal{L}\{x(t)\}(s)} = H_{00}(s) \tag{23}$$

### III. MODELING OF CIRCUITS

Power electronic circuits can be described by time-variant state-space models with switching signals as parameters. The moving Fourier coefficients of the variables of a state-space model form a time-invariant generalized average model, which can be further linearized at a steady-state operating point to obtain an LTI small-signal model.

### A. State-space model

The state-space model of a power electronic circuit can be uniformly expressed as

$$\begin{cases} E\boldsymbol{x}'(t) = \boldsymbol{f}[\boldsymbol{x}(t), \boldsymbol{u}(t), \boldsymbol{s}(t)] \\ \boldsymbol{y}(t) = \boldsymbol{g}[\boldsymbol{x}(t), \boldsymbol{u}(t), \boldsymbol{s}(t)] \end{cases} \quad (24)$$

where the bold $\boldsymbol{x}$, $\boldsymbol{u}$, $\boldsymbol{s}$, and $\boldsymbol{y}$ represent the vectors of states, inputs, switching signals, and outputs, respectively, e.g.:

$$\boldsymbol{x}(t) = \begin{bmatrix} x_1(t) \\ x_2(t) \\ \vdots \end{bmatrix} \quad (25)$$

$\boldsymbol{f}$ and $\boldsymbol{g}$ are vector-valued functions of $\boldsymbol{x}$, $\boldsymbol{u}$, and $\boldsymbol{s}$, derived from Kirchhoff's circuit laws, e.g.:

$$\boldsymbol{f}[\boldsymbol{x}(t), \boldsymbol{u}(t), \boldsymbol{s}(t)] = \begin{bmatrix} f_1[\boldsymbol{x}(t), \boldsymbol{u}(t), \boldsymbol{s}(t)] \\ f_2[\boldsymbol{x}(t), \boldsymbol{u}(t), \boldsymbol{s}(t)] \\ \vdots \end{bmatrix} \quad (26)$$

and $E$ is a coefficient matrix containing inductances, capacitances, etc.

### B. Large-signal generalized average model

In a circuit, $\boldsymbol{x}$, $\boldsymbol{y}$, $\boldsymbol{f}$, and $\boldsymbol{g}$ are all of bounded variation in a finite fundamental period. The large-signal generalized average model of the circuit can be derived from the state-space model by using the properties of moving Fourier coefficients and expressed as

$$\begin{cases} (E \otimes I)\dfrac{d\langle\boldsymbol{x}\rangle}{d\tau}(\tau) = \langle\boldsymbol{f}\rangle(\tau) - (E \otimes jN\omega)\langle\boldsymbol{x}\rangle(\tau) \\ \langle\boldsymbol{y}\rangle(\tau) = \langle\boldsymbol{g}\rangle(\tau) \end{cases} \quad (27)$$

where $\langle\boldsymbol{x}\rangle$, $\langle\boldsymbol{y}\rangle$, $\langle\boldsymbol{f}\rangle$ and $\langle\boldsymbol{g}\rangle$ are the moving Fourier coefficients of the vector-valued signals, e.g.:

$$\langle\boldsymbol{x}\rangle(\tau) = \begin{bmatrix} \langle x_1\rangle(\tau) \\ \langle x_2\rangle(\tau) \\ \vdots \end{bmatrix} \quad (28)$$

$\omega$ is the fundamental frequency, and "$\otimes$" denotes the Kronecker product, e.g.:

$$E \otimes I = \begin{bmatrix} e_{11}I & e_{12}I & \cdots \\ e_{21}I & e_{22}I & \cdots \\ \vdots & \vdots & \ddots \end{bmatrix} \quad (29)$$

where $e_{11}, e_{12}, e_{21}, e_{22}$, etc. are the elements of $E$.

### C. The steady state

In the steady state, $\boldsymbol{x}$, $\boldsymbol{u}$, $\boldsymbol{s}$, $\boldsymbol{y}$, $\boldsymbol{f}$, and $\boldsymbol{g}$ are periodical signals with the same period. If the period is taken as the fundamental period, then $\langle\boldsymbol{x}\rangle$, $\langle\boldsymbol{u}\rangle$, $\langle\boldsymbol{s}\rangle$, $\langle\boldsymbol{y}\rangle$, $\langle\boldsymbol{f}\rangle$ and $\langle\boldsymbol{g}\rangle$ are all constants. Given $\langle\boldsymbol{u}\rangle$ and $\langle\boldsymbol{s}\rangle$, $\langle\boldsymbol{f}\rangle$ and $\langle\boldsymbol{g}\rangle$ are linear functions of $\langle\boldsymbol{x}\rangle$, and the steady-state values of $\langle\boldsymbol{x}\rangle$ and $\langle\boldsymbol{y}\rangle$ can be solved from the linear equilibrium equation:

$$\begin{cases} \langle\boldsymbol{f}\rangle - (E \otimes jN\omega)\langle\boldsymbol{x}\rangle = 0 \\ \langle\boldsymbol{y}\rangle = \langle\boldsymbol{g}\rangle \end{cases} \quad (30)$$

State-dependent switching signals can also be determined by an iteration process using the linear equilibrium equation with an initial estimation of $\langle\boldsymbol{s}\rangle$.

### D. Small-signal generalized average model

The small-signal generalized average model of the circuit in the steady state can be derived from the large-signal model through local linearization and expressed as

$$\begin{cases} (E \otimes I)\dfrac{d\widetilde{\langle\boldsymbol{x}\rangle}}{d\tau}(\tau) = \left(\dfrac{\partial\langle\boldsymbol{f}\rangle}{\partial\langle\boldsymbol{x}\rangle} - E \otimes jN\omega\right)\widetilde{\langle\boldsymbol{x}\rangle}(\tau) \\ \qquad\qquad\qquad + \begin{bmatrix}\dfrac{\partial\langle\boldsymbol{f}\rangle}{\partial\langle\boldsymbol{u}\rangle} & \dfrac{\partial\langle\boldsymbol{f}\rangle}{\partial\langle\boldsymbol{s}\rangle}\end{bmatrix}\begin{bmatrix}\widetilde{\langle\boldsymbol{u}\rangle}(\tau) \\ \widetilde{\langle\boldsymbol{s}\rangle}(\tau)\end{bmatrix} \\ \widetilde{\langle\boldsymbol{y}\rangle}(\tau) = \dfrac{\partial\langle\boldsymbol{g}\rangle}{\partial\langle\boldsymbol{x}\rangle}\widetilde{\langle\boldsymbol{x}\rangle}(\tau) + \begin{bmatrix}\dfrac{\partial\langle\boldsymbol{g}\rangle}{\partial\langle\boldsymbol{u}\rangle} & \dfrac{\partial\langle\boldsymbol{g}\rangle}{\partial\langle\boldsymbol{s}\rangle}\end{bmatrix}\begin{bmatrix}\widetilde{\langle\boldsymbol{u}\rangle}(\tau) \\ \widetilde{\langle\boldsymbol{s}\rangle}(\tau)\end{bmatrix} \end{cases} \quad (31)$$

where $\widetilde{\langle\boldsymbol{x}\rangle}$, $\widetilde{\langle\boldsymbol{u}\rangle}$, $\widetilde{\langle\boldsymbol{s}\rangle}$, and $\widetilde{\langle\boldsymbol{y}\rangle}$ are the small signals of $\langle\boldsymbol{x}\rangle$, $\langle\boldsymbol{u}\rangle$, $\langle\boldsymbol{s}\rangle$, and $\langle\boldsymbol{y}\rangle$, respectively, $\partial\langle\boldsymbol{f}\rangle/\partial\langle\boldsymbol{x}\rangle$, $\partial\langle\boldsymbol{f}\rangle/\partial\langle\boldsymbol{u}\rangle$, $\partial\langle\boldsymbol{f}\rangle/\partial\langle\boldsymbol{s}\rangle$, $\partial\langle\boldsymbol{g}\rangle/\partial\langle\boldsymbol{x}\rangle$, $\partial\langle\boldsymbol{g}\rangle/\partial\langle\boldsymbol{u}\rangle$, and $\partial\langle\boldsymbol{g}\rangle/\partial\langle\boldsymbol{s}\rangle$ are Jacobian matrices in the form as

$$\dfrac{\partial\langle\boldsymbol{f}\rangle}{\partial\langle\boldsymbol{x}\rangle} = \begin{bmatrix} \dfrac{\partial\langle f_1\rangle}{\partial\langle x_1\rangle} & \dfrac{\partial\langle f_1\rangle}{\partial\langle x_2\rangle} & \cdots \\ \dfrac{\partial\langle f_2\rangle}{\partial\langle x_1\rangle} & \dfrac{\partial\langle f_2\rangle}{\partial\langle x_2\rangle} & \cdots \\ \vdots & \vdots & \ddots \end{bmatrix} = \begin{bmatrix} \left[\!\!\left[\dfrac{\partial f_1}{\partial x_1}\right]\!\!\right] & \left[\!\!\left[\dfrac{\partial f_1}{\partial x_2}\right]\!\!\right] & \cdots \\ \left[\!\!\left[\dfrac{\partial f_2}{\partial x_1}\right]\!\!\right] & \left[\!\!\left[\dfrac{\partial f_2}{\partial x_2}\right]\!\!\right] & \cdots \\ \vdots & \vdots & \ddots \end{bmatrix} \quad (32)$$

where each element of the Jacobian matrix is a Toeplitz matrix of moving Fourier coefficients of a partial derivative determined by the steady-state values.

## IV. MODELING OF SWITCHING SIGNALS

Switching signals are Boolean signals determined by their rising and falling instants. The rising and falling instants are usually determined by the zero points of other signals. The small-signal generalized average model of a switching signal is the combination of the partial derivatives of its moving Fourier coefficients with respect to its rising and falling instants and the partial derivatives of the zero points with respect to the moving Fourier coefficients of the related signals.

### A. Rising and falling instants

If a switching signal $s(t)$ on $[\tau - T/2, \tau + T/2]$ has a unique rising instant $t_r(\tau)$ and a unique falling instant $t_f(\tau)$, as shown in Fig. 4, then its moving Fourier coefficients at $\tau$ are functions of $t_r(\tau)$ and $t_f(\tau)$ as

$$\begin{cases} \langle s\rangle_0(\tau) = \mathrm{mod}\left[\dfrac{t_f(\tau) - t_r(\tau)}{T}, 1\right] \\ \angle\langle s\rangle_1(\tau) = \mathrm{wrap}[-\omega t_r(\tau) - \pi\langle s\rangle_0(\tau)] \\ \qquad\quad\;\; = \mathrm{wrap}[-\omega t_f(\tau) + \pi\langle s\rangle_0(\tau)] \end{cases} \quad (33)$$

and for any $n \in \mathbb{Z}$,

$$\langle s\rangle_n(\tau) = \begin{cases} \langle s\rangle_0(\tau), & n = 0 \\ \dfrac{1}{\pi n}e^{jn\angle\langle s\rangle_1(\tau)}\sin\pi n\langle s\rangle_0(\tau), & n \neq 0 \end{cases} \quad (34)$$

where "mod" is the remainder function, "∠" denotes the angle of a complex number, and "wrap" wraps an angle to $(-\pi, \pi]$.

Fig. 4. Switching signal.

The partial derivatives of $\langle s \rangle_n$ with respect to $t_r(\tau)$ and $t_f(\tau)$ are

$$\begin{cases} \dfrac{\partial \langle s \rangle_n}{\partial t_r}(\tau) = -\dfrac{1}{T}\mathrm{e}^{-jn\omega t_r(\tau)} \\ \dfrac{\partial \langle s \rangle_n}{\partial t_f}(\tau) = \dfrac{1}{T}\mathrm{e}^{-jn\omega t_f(\tau)} \end{cases} \quad (35)$$

In vector form:

$$\frac{\partial \langle s \rangle}{\partial t_r}(\tau) = \begin{bmatrix} \vdots \\ \dfrac{\partial \langle s \rangle_{-1}}{\partial t_r}(\tau) \\ \dfrac{\partial \langle s \rangle_0}{\partial t_r}(\tau) \\ \dfrac{\partial \langle s \rangle_1}{\partial t_r}(\tau) \\ \vdots \end{bmatrix}, \quad \frac{\partial \langle s \rangle}{\partial t_f}(\tau) = \begin{bmatrix} \vdots \\ \dfrac{\partial \langle s \rangle_{-1}}{\partial t_f}(\tau) \\ \dfrac{\partial \langle s \rangle_0}{\partial t_f}(\tau) \\ \dfrac{\partial \langle s \rangle_1}{\partial t_f}(\tau) \\ \vdots \end{bmatrix} \quad (36)$$

where $t_r(\tau)$ and $t_f(\tau)$ can also be expressed by $\langle s \rangle_0(\tau)$ and $\angle \langle s \rangle_1(\tau)$ as

$$\begin{cases} t_r(\tau) = \tau + \dfrac{1}{\omega}\mathrm{wrap}[-\omega\tau - \angle\langle s \rangle_1(\tau) - \pi\langle s \rangle_0(\tau)] \\ t_f(\tau) = \tau + \dfrac{1}{\omega}\mathrm{wrap}[-\omega\tau - \angle\langle s \rangle_1(\tau) + \pi\langle s \rangle_0(\tau)] \end{cases} \quad (37)$$

Since $t_r(\tau)$ and $t_f(\tau)$ only appear as powers in the exponential functions of (35), they can be equivalent to

$$\begin{cases} t_r(\tau) \equiv \dfrac{1}{\omega}\mathrm{wrap}[-\angle\langle s \rangle_1(\tau) - \pi\langle s \rangle_0(\tau)] \\ t_f(\tau) \equiv \dfrac{1}{\omega}\mathrm{wrap}[-\angle\langle s \rangle_1(\tau) + \pi\langle s \rangle_0(\tau)] \end{cases} \quad (38)$$

In the steady state, $s(t)$ is a periodic signal, and the distance between each two rising (or falling) instants is the fundamental period $T$. Therefore, $s(t)$ has a unique rising instant and a unique falling instant on $[\tau - T/2, \tau + T/2)$ for any $\tau$, and (35) is valid almost everywhere, i.e., the set of $\tau$ for which (35) is not valid is of measure zero.

B. *Zero point of signal*

If $x(t)$ and $x'(t)$ are of bounded variation on $[\tau - T/2, \tau + T/2]$, and $x(t)$ crosses, exceeds or decreases to zero at $t_0(\tau)$, then

$$x[t_0(\tau)] = \sum_{n=-\infty}^{+\infty} \langle x \rangle_n(\tau)\mathrm{e}^{jn\omega t_0(\tau)} = 0 \quad (39)$$

By taking the full differential, it yields

$$\sum_{n=-\infty}^{+\infty} \mathrm{e}^{jn\omega t_0(\tau)} \mathrm{d}\langle x \rangle_n(\tau) + \sum_{n=-\infty}^{+\infty} jn\omega\langle x \rangle_n(\tau)\mathrm{e}^{jn\omega t_0(\tau)} \mathrm{d}t_0(\tau) = 0 \quad (40)$$

Consequently, for any $m \in \mathbb{Z}$,

$$\frac{\partial t_0}{\partial \langle x \rangle_m}(\tau) = \frac{-\mathrm{e}^{jm\omega t_0(\tau)}}{\sum_{n=-\infty}^{+\infty} jn\omega\langle x \rangle_n(\tau)\mathrm{e}^{jn\omega t_0(\tau)}} \quad (41)$$

In vector form:

$$\frac{\partial t_0}{\partial \langle x \rangle}(\tau) = \begin{bmatrix} \cdots & \dfrac{\partial t_0}{\partial \langle x \rangle_{-1}}(\tau) & \dfrac{\partial t_0}{\partial \langle x \rangle_0}(\tau) & \dfrac{\partial t_0}{\partial \langle x \rangle_1}(\tau) & \cdots \end{bmatrix} \quad (42)$$

In addition, according to the property of time derivative, (41) is rewritten as

$$\frac{\partial t_0}{\partial \langle x \rangle_m}(\tau) = \frac{-\mathrm{e}^{jm\omega t_0(\tau)}}{\lim\limits_{\varepsilon \to 0} \dfrac{x'[t_0(\tau) + \varepsilon] + x'[t_0(\tau) - \varepsilon]}{2}} \quad (43)$$

In particular, if $x(t)$ is differentiable at $t_0(\tau)$, then

$$\frac{\partial t_0}{\partial \langle x \rangle_m}(\tau) = \frac{-\mathrm{e}^{jm\omega t_0(\tau)}}{x'[t_0(\tau)]} \quad (44)$$

The expression of (41) is a general small-signal relationship between the zero point and the moving Fourier coefficients of a signal, while (43) and (44) are useful alternatives in special cases. For example, the turn-off instant of a diode is the zero point of its forward current. In DCM, the forward current decreases with a certain slope to zero and then remains unchanged. The current is not differentiable at the zero point, but its time derivative has left and right limits, i.e., the decreasing slope and zero, which can be used in (43). The combination of (41) or (43) and (35) gives the small-signal generalized average models of state-dependent switching signals such as the switching signal of the diode. On the other hand, a representative application of (44) is the PWM signal that rises and falls at the zero points of the difference between the duty-cycle $d(t)$ and the carrier $c(t)$, i.e. $e(t) = d(t) - c(t)$. With small-signal perturbations, $e'(t) = -c'(t)$, and according to (44), there is

$$\frac{\partial t_0}{\partial \langle d \rangle_m}(\tau) = \frac{\partial t_0}{\partial \langle e \rangle_m}(\tau) = \frac{-\mathrm{e}^{jm\omega t_0(\tau)}}{e'[t_0(\tau)]} = \frac{\mathrm{e}^{jm\omega t_0(\tau)}}{c'[t_0(\tau)]} \quad (45)$$

This equation applies to PWM with any carrier, including triangle carrier, sawtooth carrier, reverse sawtooth carrier, etc., where $c'[t_0(\tau)]$ is determined by the shape of the carrier (see Table 2). The combination of (45) and (35) gives the small-signal generalized average model of PWM signals.

C. *Zero point of function of signal*

If $x(t)$, $f(x,t)$, and $\partial f/\partial x$, are all of bounded variation on $[\tau - T/2, \tau + T/2]$, and $f(x,t)$ crosses, exceeds, or decreases to zero at $t_0(\tau)$, then, by referring to (40), there is

$$\sum_{n=-\infty}^{+\infty} \mathrm{e}^{jn\omega t_0(\tau)} \mathrm{d}\langle f \rangle_n(\tau) + \sum_{n=-\infty}^{+\infty} jn\omega\langle f \rangle_n(\tau)\mathrm{e}^{jn\omega t_0(\tau)} \mathrm{d}t_0(\tau) = 0 \quad (46)$$

If $x(t)$ and $f(x,t)$ are differentiable at $t_0(\tau)$, then

$$\sum_{n=-\infty}^{+\infty} \mathrm{e}^{jn\omega t_0(\tau)} \mathrm{d}\langle f \rangle_n(\tau) = \mathrm{d}f[t_0(\tau)]$$

$$= \frac{\partial f}{\partial x}[t_0(\tau)]\mathrm{d}x[t_0(\tau)]$$

$$= \frac{\partial f}{\partial x}[t_0(\tau)] \sum_{n=-\infty}^{+\infty} \mathrm{e}^{jn\omega t_0(\tau)} \mathrm{d}\langle x \rangle_n(\tau) \quad (47)$$

By substituting (47) into (46), it yields

$$\frac{\partial t_0}{\partial \langle x \rangle_m}(\tau) = \frac{\partial f}{\partial x}[t_0(\tau)] \frac{-\mathrm{e}^{jm\omega t_0(\tau)}}{\sum_{n=-\infty}^{+\infty} jn\omega\langle f \rangle_n(\tau)\mathrm{e}^{jn\omega t_0(\tau)}}$$

$$= \frac{\partial f}{\partial x}[t_0(\tau)] \frac{-\mathrm{e}^{jm\omega t_0(\tau)}}{f'[t_0(\tau)]} \quad (48)$$

for any $m \in \mathbb{Z}$. In vector form:

$$\frac{\partial t_0}{\partial \langle x \rangle}(\tau) = \frac{\partial f}{\partial x}[t_0(\tau)] \frac{\partial t_0}{\partial \langle f \rangle}(\tau) \tag{49}$$

For example, the rising and falling instants of a PSM signal are the zero points of the modulated carrier, which is a function of $t$ and the phase-shift signal $\alpha(t)$ as $c[\omega t + \alpha(t)]$. With small-signal perturbations, $\partial c/\partial \alpha = c'/\omega$, and according to (48), there is

$$\frac{\partial t_0}{\partial \langle \alpha \rangle_m}(\tau) = \frac{\partial c}{\partial \alpha}[t_0(\tau)] \frac{-e^{jm\omega t_0(\tau)}}{c'[t_0(\tau)]} = \frac{-e^{jm\omega t_0(\tau)}}{\omega} \tag{50}$$

This equation applies to PSM with any carrier, including sine carrier, triangle carrier, sawtooth carrier, reverse sawtooth carrier, etc., and is independent of the shape of the carrier. The combination of (50) and (35) gives the small-signal generalized average model of PSM signals.

Furthermore, if using the time derivative of $\alpha(t)$ to represent the frequency of a PFM signal as

$$\omega_s(t) = \omega + \alpha'(t) \tag{51}$$

then the transfer function from $\omega_s(t)$ to any signal of interest, e.g., $y(t)$, can be derived from the transfer function from $\alpha(t)$ to the signal by multiplying it by $1/s$, i.e.

$$\frac{\mathcal{L}\{y(t)\}(s)}{\mathcal{L}\{\omega_s(t)\}(s)} = \frac{1}{s} \cdot \frac{\mathcal{L}\{y(t)\}(s)}{\mathcal{L}\{\alpha(t)\}(s)} \tag{52}$$

In this way, the modeling of PFM signals translates to the modeling of PSM signals.

## V. PWM CONTROLLED BOOST CONVERTER

This section presents the modeling process of a typical PWM controlled boost converter (see Fig. 5) as a basic example of the generalized averaging method.

### A. Generalized average model of the circuit

The state-space model of the circuit is

$$\begin{bmatrix} L & 0 \\ 0 & C \end{bmatrix} \frac{d}{dt} \begin{bmatrix} i_L \\ v_o \end{bmatrix} = \begin{bmatrix} (s_1 + s_2)v_i - s_2 v_o \\ s_2 i_L - \frac{1}{R} v_o \end{bmatrix} \tag{53}$$

where $L$, $C$, and $R$ are the choke inductance, filter capacitance, and load resistance, respectively, $i_L$, $v_i$ and $v_o$ are the inductor current, input voltage, and output voltage, respectively, $s_1$ and $s_2$ are the switching signals of switch S and diode D, respectively.

According to (27), the large-signal generalized average model of the circuit is derived from (53) and expressed as

$$\begin{bmatrix} LI & 0 \\ 0 & CI \end{bmatrix} \frac{d}{d\tau} \begin{bmatrix} \langle i_L \rangle \\ \langle v_o \rangle \end{bmatrix}$$
$$= \begin{bmatrix} (\langle s_1 \rangle + \langle s_2 \rangle) * \langle v_i \rangle - \langle s_2 \rangle * \langle v_o \rangle - jN\omega L \langle i_L \rangle \\ \langle s_2 \rangle * \langle i_L \rangle - \frac{1}{R} \langle v_o \rangle - jN\omega C \langle v_o \rangle \end{bmatrix} \tag{54}$$

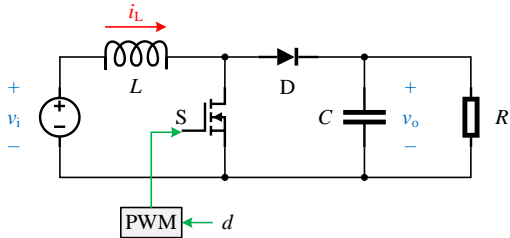

Fig. 5. PWM controlled boost converter.

According to (31), the small-signal generalized average model is derived from (54) and expressed as

$$\begin{bmatrix} LI & 0 \\ 0 & CI \end{bmatrix} \frac{d}{d\tau} \begin{bmatrix} \widetilde{\langle i_L \rangle} \\ \widetilde{\langle v_o \rangle} \end{bmatrix} = \begin{bmatrix} -jN\omega L & -[\![s_2]\!] \\ [\![s_2]\!] & -\frac{I}{R} - jN\omega C \end{bmatrix} \begin{bmatrix} \widetilde{\langle i_L \rangle} \\ \widetilde{\langle v_o \rangle} \end{bmatrix}$$
$$+ \begin{bmatrix} [\![v_i]\!] & [\![v_i]\!] - [\![v_o]\!] \\ 0 & [\![i_L]\!] \end{bmatrix} \begin{bmatrix} \widetilde{\langle s_1 \rangle} \\ \widetilde{\langle s_2 \rangle} \end{bmatrix} \tag{55}$$

### B. Generalized average model of the switching signals

For PWM control, the rising and falling instants of $s_1$, i.e. $t_{r1}$ and $t_{f1}$, are the zero points of $d - c$, where $d$ is the duty cycle and $c$ is the carrier. Therefore, the small-signal generalized average model of $s_1$ is the combination of

$$\widetilde{\langle s_1 \rangle} = \begin{bmatrix} \frac{\partial \langle s_1 \rangle}{\partial t_{r1}} & \frac{\partial \langle s_1 \rangle}{\partial t_{f1}} \end{bmatrix} \begin{bmatrix} \widetilde{t_{r1}} \\ \widetilde{t_{f1}} \end{bmatrix} \tag{56}$$

and

$$\begin{bmatrix} \widetilde{t_{r1}} \\ \widetilde{t_{f1}} \end{bmatrix} = \begin{bmatrix} \frac{\partial t_{r1}}{\partial \langle d \rangle} \\ \frac{\partial t_{f1}}{\partial \langle d \rangle} \end{bmatrix} \widetilde{\langle d \rangle} \tag{57}$$

According to (35), for any $n \in \mathbb{Z}$

$$\begin{cases} \frac{\partial \langle s_1 \rangle_n}{\partial t_{r1}} = -\frac{1}{T} e^{-jn\omega t_{r1}} \\ \frac{\partial \langle s_1 \rangle_n}{\partial t_{f1}} = \frac{1}{T} e^{-jn\omega t_{f1}} \end{cases} \tag{58}$$

According to (45), for any $m \in \mathbb{Z}$

$$\begin{cases} \frac{\partial t_{r1}}{\partial \langle d \rangle_m} = \frac{e^{jm\omega t_{r1}}}{c'(t_{r1})} \\ \frac{\partial t_{f1}}{\partial \langle d \rangle_m} = \frac{e^{jm\omega t_{f1}}}{c'(t_{f1})} \end{cases} \tag{59}$$

Table 2 shows the specific model of $s_1$ for different carriers.

In CCM, $s_2$ is the complementary signal of $s_1$, and therefore, the small-signal generalized average model of $s_2$ is

$$\widetilde{\langle s_2 \rangle} = -\widetilde{\langle s_1 \rangle} \tag{60}$$

In DCM, the rising instant $t_{r2}$ of $s_2$ is the falling instant of $s_1$, i.e. $t_{f1}$, the falling instant $t_{f2}$ of $s_2$ is the zero point of $i_L$, and therefore, the small-signal generalized average model of $s_2$ is the combination of

$$\widetilde{\langle s_2 \rangle} = \begin{bmatrix} \frac{\partial \langle s_2 \rangle}{\partial t_{r2}} & \frac{\partial \langle s_2 \rangle}{\partial t_{f2}} \end{bmatrix} \begin{bmatrix} \widetilde{t_{r2}} \\ \widetilde{t_{f2}} \end{bmatrix} \tag{61}$$

and

$$\begin{bmatrix} \widetilde{t_{r2}} \\ \widetilde{t_{f2}} \end{bmatrix} = \begin{bmatrix} 1 & \langle 0 \rangle^{\mathrm{T}} \\ 0 & \frac{\partial t_{f2}}{\partial \langle i_L \rangle} \end{bmatrix} \begin{bmatrix} \widetilde{t_{f1}} \\ \widetilde{\langle i_L \rangle} \end{bmatrix} \tag{62}$$

where $\langle 0 \rangle$ is a zero vector and the superscript "T" indicates transpose. According to (35), for any $n \in \mathbb{Z}$

$$\begin{cases} \frac{\partial \langle s_2 \rangle_n}{\partial t_{r2}} = -\frac{1}{T} e^{-jn\omega t_{r2}} \\ \frac{\partial \langle s_2 \rangle_n}{\partial t_{f2}} = \frac{1}{T} e^{-jn\omega t_{f2}} \end{cases} \tag{63}$$

According to (41), for any $m \in \mathbb{Z}$

$$\frac{\partial t_{f2}}{\partial \langle i_L \rangle_m} = \frac{-e^{jm\omega t_{f2}}}{\sum_{n=-\infty}^{+\infty} jn\omega \langle i_L \rangle_n e^{jn\omega t_{f2}}} \tag{64}$$

Table 2. PWM signal models with different carriers

| Triangle carrier | Sawtooth carrier | Reverse sawtooth carrier |
|---|---|---|
| 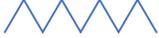 | 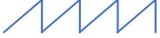 | 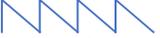 |
| $c(t) = \dfrac{1}{\pi}\|\mathrm{wrap}(\omega t)\|$ | $c(t) \to 0.5 + \dfrac{1}{2\pi}\mathrm{wrap}(\omega t)$ | $c(t) \to 0.5 - \dfrac{1}{2\pi}\mathrm{wrap}(\omega t)$ |
| $\begin{cases} c'(t_{r1}) = -\dfrac{2}{T} \\ c'(t_{f1}) = \dfrac{2}{T} \end{cases}$ | $\begin{cases} c'(t_{r1}) = -\infty \\ c'(t_{f1}) = \dfrac{1}{T} \end{cases}$ | $\begin{cases} c'(t_{r1}) = -\dfrac{1}{T} \\ c'(t_{f1}) = +\infty \end{cases}$ |
| $\dfrac{\partial \langle s_1 \rangle_n}{\partial \langle d \rangle_m} = \dfrac{e^{-j(n-m)\omega t_{r1}} + e^{-j(n-m)\omega t_{f1}}}{2}$ | $\dfrac{\partial \langle s_1 \rangle_n}{\partial \langle d \rangle_m} = e^{-j(n-m)\omega t_{f1}}$ | $\dfrac{\partial \langle s_1 \rangle_n}{\partial \langle d \rangle_m} = e^{-j(n-m)\omega t_{r1}}$ |

### C. Steady-state operating point

According to (34), $\langle s_1 \rangle$ and $\langle s_2 \rangle$ are expressed as

$$\langle s_1 \rangle_n = \begin{cases} \langle s_1 \rangle_0, & n = 0 \\ \dfrac{1}{\pi n} e^{jn\angle\langle s_1 \rangle_1} \sin \pi n \langle s_1 \rangle_0, & n \neq 0 \end{cases} \quad (65)$$

and

$$\langle s_2 \rangle_n = \begin{cases} \langle s_2 \rangle_0, & n = 0 \\ \dfrac{1}{\pi n} e^{jn\angle\langle s_2 \rangle_1} \sin \pi n \langle s_2 \rangle_0, & n \neq 0 \end{cases} \quad (66)$$

In the steady state, $\langle s_1 \rangle_0$ is $d$, $\angle\langle s_1 \rangle_1$ is 0 if taking $s_1$ as the phase reference, and $\angle\langle s_2 \rangle_1$ is $-\pi\langle s_1 \rangle_0 - \pi\langle s_2 \rangle_0$ because $t_{r2}$ equals $t_{f1}$. In CCM, $\langle s_2 \rangle_0$ is $1-d$. In DCM, $\langle s_2 \rangle_0$ can be determined by iterating $i_L(t_{f2})$ towards 0.

Given $\langle v_i \rangle$, $\langle s_1 \rangle$ and $\langle s_2 \rangle$, the steady-state values of $\langle i_L \rangle$ and $\langle v_o \rangle$ can be solved from the linear equilibrium equation:

$$\begin{bmatrix} (\llbracket s_1 \rrbracket + \llbracket s_2 \rrbracket)\langle v_i \rangle - \llbracket s_2 \rrbracket \langle v_o \rangle - jN\omega L \langle i_L \rangle \\ \llbracket s_2 \rrbracket \langle i_L \rangle - \dfrac{1}{R}\langle v_o \rangle - jN\omega C \langle v_o \rangle \end{bmatrix} = 0 \quad (67)$$

and $i_L(t_{f2})$ can be calculated using $\langle i_L \rangle$ according to (3) and (38) in the iteration.

### D. Experiment

The boost converter prototype used for the experiments is shown in Fig. 6. The parameters are shown in Table 3. The converter was controlled by a PWM signal generated by a signal generator. When the load resistance was 25 Ω, the circuit worked in CCM and the output voltage was about 48 V. When the load resistance was 100 Ω, the circuit worked in DCM and the output voltage was about 60 V. The capacitance of the filter ceramic capacitors had different attenuation at different DC bias voltages. Based on the output voltage ripples measured in the experiments, the filter capacitance was estimated to be 0.9 μF at 48 V bias voltage in CCM and 0.7 μF at 60 V bias voltage in DCM.

Fig. 7 (a) and (b) show the steady-state waveforms captured in CCM and DCM respectively. Fig. 7 (c) and (d) show the waveforms derived from the corresponding steady-state solutions for the −49th to 49th moving Fourier coefficients. The steady-state solutions were obtained through several iterations of the linear equilibrium equation (67) and were consistent with the captured waveforms since sufficient harmonics were taken into account.

To obtain the frequency-domain characteristics, the duty cycle of the PWM signal was modulated with a modulation depth of 1% and a modulation frequency ranging from 1 kHz to 180 kHz, and the responses of the converter were recorded. For example, Fig. 7 (e) and (f) show the responses captured at a modulation frequency of 10 kHz in CCM and 150 kHz in DCM, respectively. From the spectrum of the response data, the value of the duty cycle-to-output voltage transfer function at each modulation frequency was obtained, as shown in Fig. 7 (g) and (h). For comparison, Fig. 7 (g) and (h) also show the transfer function curves predicted by the generalized average model built in this section and the classical moving average model [5, 30]. The generalized average model is more accurate than the moving average model when the modulation frequency is close to or higher than half the switching frequency.

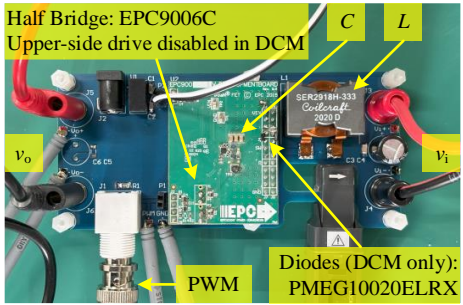

Fig. 6. Boost converter prototype.

Table 3. Boost converter parameters

| Symbol | Quantity | Value | |
|---|---|---|---|
| | | CCM | DCM |
| $L$ | Choke inductance | 33 μH | 33 μH |
| $C$ | Filter capacitance | 0.9 μF | 0.7 μF |
| $R$ | Load resistance | 25 Ω | 100 Ω |
| $v_i$ | Input voltage | 24 V | 24 V |
| $f_s$ | Switching frequency | 100 kHz | 100 kHz |
| $d$ | Steady-state duty cycle | 50% | 50% |

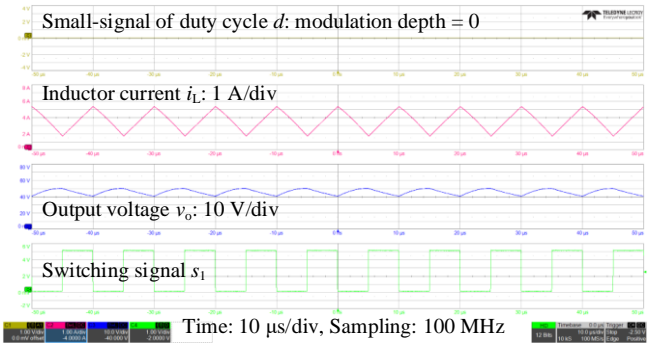

(a)

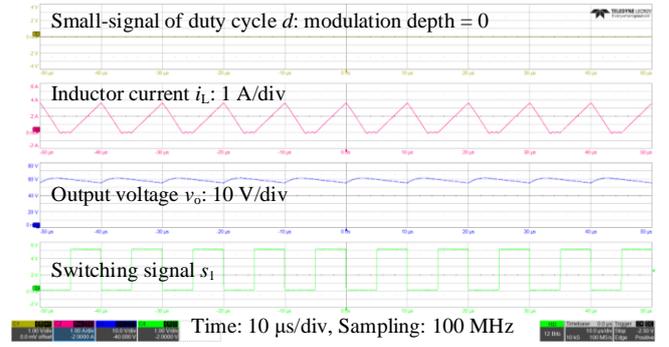

(b)

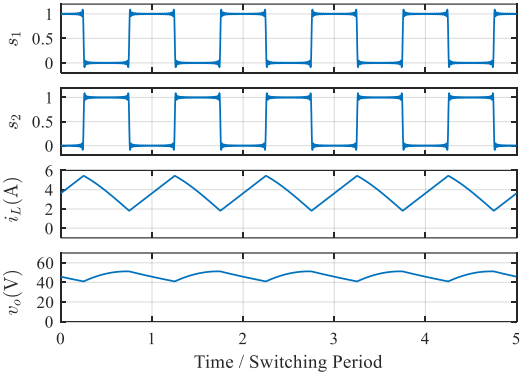

(c)

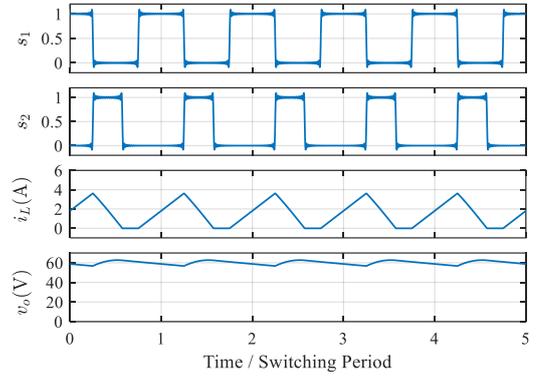

(d)

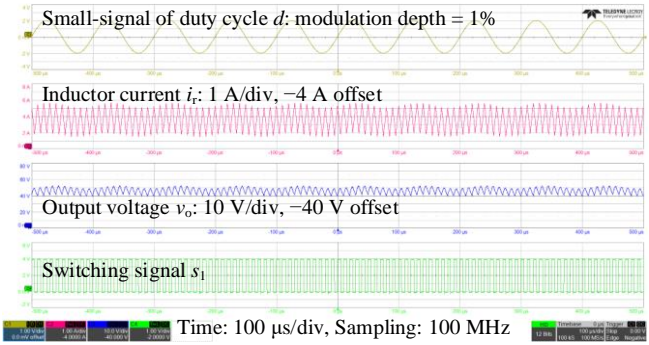

(e)

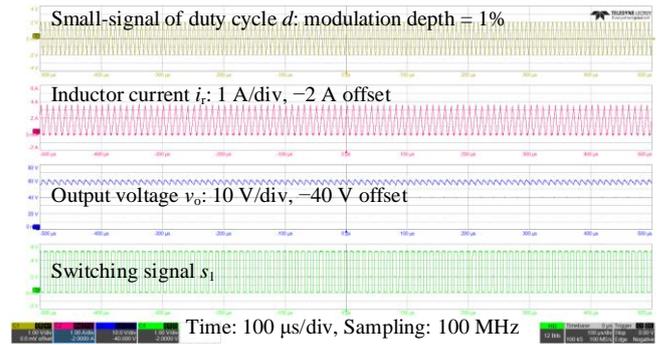

(f)

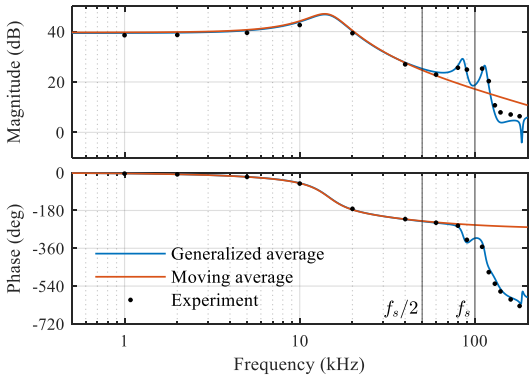

(g)

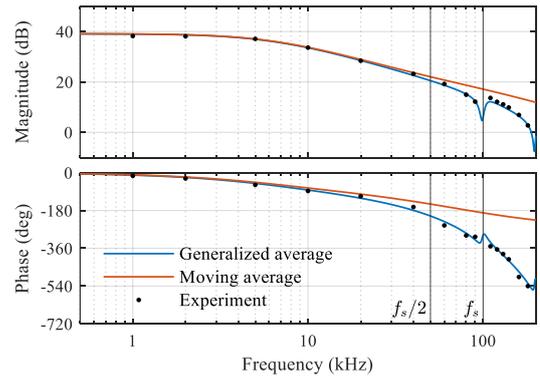

(h)

Fig. 7. Experimental and calculation results of the boost converter: steady-state waveforms captured in (a) CCM and (b) DCM, waveforms derived from the steady-state solutions for the moving Fourier coefficients in (c) CCM and (d) DCM, responses captured when the duty cycle was modulated with a modulation depth of 1% and a modulation frequency of (e) 10 kHz in CCM and (f) 150 kHz in DCM, duty cycle-to-out voltage transfer function in (g) CCM and (h) DCM.

## VI. V² CONSTANT ON-TIME CONTROLLED BUCK CONVERTER

Current-mode controlled buck converters have long been a challenging subject in high-frequency modeling. This type of converter has fast dynamic response and inherent over-current protection, and is widely used in point-of-load applications to meet load-line requirements [31]. In current-mode control, the sensed inductor current is used as the PWM ramp without any low-pass filter. Therefore, the high-frequency information is crucial for the modeling. Modulation schemes for current-mode control include peak current-mode control, valley current-mode control, constant on-time control, constant off-time control, etc. [32]. The constant on/off-time control belongs to variable-frequency modulation. In practical designs, the V² implementation of constant on-time control has gained more attention due to its high light-load efficiency and simplicity. The circuit is shown in Fig. 8, where the resistance $r$ (possibly from the equivalent series resistance of the filter capacitor) senses the inductor current, so the output voltage ripple can be directly used as the PWM ramp [14].

To date, one of the most accurate models of V² constant on-time controlled buck converters is J. Li and F. C. Lee's describing function model [15]. The modeling process relies on the detailed waveform analysis. In contrast, the generalized average model built in this section does not require delving into the details of the waveforms, and has higher accuracy at high frequencies. In addition to CCM, the model of the converter in DCM is also built for the first time.

### A. Generalized average model of the circuit

The state-space model of the circuit is
$$\begin{cases} \begin{bmatrix} L & 0 \\ 0 & C \end{bmatrix} \frac{d}{dt} \begin{bmatrix} i_L \\ u_C \end{bmatrix} = \begin{bmatrix} s_1 v_i - (s_1 + s_2) v_o \\ i_L - v_o/R \end{bmatrix} \\ v_o = \frac{Rr}{R+r} i_L + \frac{R}{R+r} u_C \end{cases} \quad (68)$$

where $L$, $C$, and $R$ are the choke inductance, filter capacitance, and load resistance, respectively, $i_L$, $u_C$, $v_i$, and $v_o$ are the inductor current, capacitor voltage, input voltage, and output voltage, respectively, $s_1$ and $s_2$ are the switching signals of switch S and diode D, respectively.

According to (27), the large-signal generalized average model of the circuit is derived from (68) and expressed as
$$\begin{cases} \begin{bmatrix} LI & 0 \\ 0 & CI \end{bmatrix} \frac{d}{d\tau} \begin{bmatrix} \langle i_L \rangle \\ \langle u_C \rangle \end{bmatrix} = \\ \quad \begin{bmatrix} \langle s_1 \rangle * \langle v_i \rangle - (\langle s_1 \rangle + \langle s_2 \rangle) * \langle v_o \rangle - jN\omega L \langle i_L \rangle \\ \langle i_L \rangle - \langle v_o \rangle / R - jN\omega C \langle u_C \rangle \end{bmatrix} \\ \langle v_o \rangle = \frac{Rr}{R+r} \langle i_L \rangle + \frac{R}{R+r} \langle u_C \rangle \end{cases} \quad (69)$$

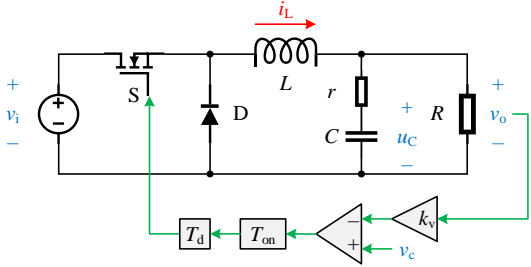

Fig. 8. V² constant on-time controlled buck converter.

According to (31), the small-signal generalized average model is derived from (69) and expressed as
$$\begin{cases} \begin{bmatrix} LI & 0 \\ 0 & CI \end{bmatrix} \frac{d}{d\tau} \begin{bmatrix} \widetilde{\langle i_L \rangle} \\ \widetilde{\langle u_C \rangle} \end{bmatrix} = \\ \begin{bmatrix} -\frac{Rr(\llbracket s_1 \rrbracket + \llbracket s_2 \rrbracket)}{R+r} - jN\omega L & -\frac{R(\llbracket s_1 \rrbracket + \llbracket s_2 \rrbracket)}{R+r} \\ \frac{RI}{R+r} & -\frac{I}{R+r} - jN\omega C \end{bmatrix} \begin{bmatrix} \widetilde{\langle i_L \rangle} \\ \widetilde{\langle u_C \rangle} \end{bmatrix} \\ + \begin{bmatrix} \llbracket v_i \rrbracket - \llbracket v_o \rrbracket & -\llbracket v_o \rrbracket \\ 0 & 0 \end{bmatrix} \begin{bmatrix} \widetilde{\langle s_1 \rangle} \\ \widetilde{\langle s_2 \rangle} \end{bmatrix} \\ \widetilde{\langle v_o \rangle} = \begin{bmatrix} \frac{Rr}{R+r} & \frac{R}{R+r} \end{bmatrix} \begin{bmatrix} \widetilde{\langle i_L \rangle} \\ \widetilde{\langle u_C \rangle} \end{bmatrix} \end{cases} \quad (70)$$

### B. Generalized average model of the switching signals

For V² constant on-time control, the rising instant $t_{r1}$ of $s_1$ is the zero point of $v_c - k_v v_o$, where $v_c$ is the control voltage and $k_v$ is the voltage detection gain. The falling instant $t_{f1}$ of $s_1$ is a delay of $t_{r1}$, and the delay time is the constant on-time $T_{on}$. In practice, there may be a non-negligible additional delay $T_d$ on both $t_{r1}$ and $t_{f1}$ due to the control logic and gate drive. Therefore, the small-signal generalized average model of $s_1$ is the combination of

$$\widetilde{\langle s_1 \rangle} = \begin{bmatrix} \frac{\partial \langle s_1 \rangle}{\partial t_{r1}} & \frac{\partial \langle s_1 \rangle}{\partial t_{f1}} \end{bmatrix} \begin{bmatrix} \widetilde{t_{r1}} \\ \widetilde{t_{f1}} \end{bmatrix} \quad (71)$$

and

$$\begin{bmatrix} \widetilde{t_{r1}} \\ \widetilde{t_{f1}} \end{bmatrix} = \begin{bmatrix} 1 \\ 1 \end{bmatrix} \begin{bmatrix} \frac{\partial t_0}{\partial \langle v_o \rangle} & \frac{\partial t_0}{\partial \langle v_c \rangle} \end{bmatrix} \begin{bmatrix} \widetilde{\langle v_o \rangle} \\ \widetilde{\langle v_c \rangle} \end{bmatrix},$$
$$\text{Output Delay} = \begin{bmatrix} T_d \\ T_{on} + T_d \end{bmatrix} \quad (72)$$

where $t_0$ is the zero point of $v_c - k_v v_o$. According to (35), for any $n \in \mathbb{Z}$,
$$\begin{cases} \frac{\partial \langle s_1 \rangle_n}{\partial t_{r1}} = -\frac{1}{T} e^{-jn\omega t_{r1}} \\ \frac{\partial \langle s_1 \rangle_n}{\partial t_{f1}} = \frac{1}{T} e^{-jn\omega t_{f1}} \end{cases} \quad (73)$$

According to (41) and (48), for any $m \in \mathbb{Z}$,
$$\begin{cases} \frac{\partial t_0}{\partial \langle v_o \rangle_m} = \frac{k_v e^{jm\omega t_0}}{\sum_{n=-\infty}^{+\infty} jn\omega \langle v_c - k_v v_o \rangle_n e^{jn\omega t_0}} \\ \frac{\partial t_0}{\partial \langle v_c \rangle_m} = \frac{-e^{jm\omega t_0}}{\sum_{n=-\infty}^{+\infty} jn\omega \langle v_c - k_v v_o \rangle_n e^{jn\omega t_0}} \end{cases} \quad (74)$$

In CCM, $s_2$ is the complementary signal of $s_1$, and therefore, the small-signal generalized average model of $s_2$ is
$$\widetilde{\langle s_2 \rangle} = -\widetilde{\langle s_1 \rangle} \quad (75)$$

In DCM, the rising instant $t_{r2}$ of $s_2$ is the falling instant of $s_1$, i.e. $t_{f1}$, the falling instant $t_{f2}$ of $s_2$ is the zero point of $i_L$, and therefore, the small-signal generalized average model of $s_2$ is the combination of

$$\widetilde{\langle s_2 \rangle} = \begin{bmatrix} \frac{\partial \langle s_2 \rangle}{\partial t_{r2}} & \frac{\partial \langle s_2 \rangle}{\partial t_{f2}} \end{bmatrix} \begin{bmatrix} \widetilde{t_{r2}} \\ \widetilde{t_{f2}} \end{bmatrix} \quad (76)$$

and

$$\begin{bmatrix} \widetilde{t_{r2}} \\ \widetilde{t_{f2}} \end{bmatrix} = \begin{bmatrix} 1 & \langle 0 \rangle^T \\ 0 & \frac{\partial t_{f2}}{\partial \langle i_L \rangle} \end{bmatrix} \begin{bmatrix} \widetilde{t_{f1}} \\ \widetilde{\langle i_L \rangle} \end{bmatrix} \quad (77)$$

According to (35), for any $n \in \mathbb{Z}$,
$$\begin{cases} \dfrac{\partial \langle s_2 \rangle_n}{\partial t_{r2}} = -\dfrac{1}{T} e^{-jn\omega t_{r2}} \\ \dfrac{\partial \langle s_2 \rangle_n}{\partial t_{f2}} = \dfrac{1}{T} e^{-jn\omega t_{f2}} \end{cases} \quad (78)$$

According to (41), for any $m \in \mathbb{Z}$,
$$\frac{\partial t_{f2}}{\partial \langle i_L \rangle_m} = \frac{-e^{jm\omega t_{f2}}}{\sum_{n=-\infty}^{+\infty} jn\omega \langle i_L \rangle_n e^{jn\omega t_{f2}}} \quad (79)$$

### C. Steady-state operating point

According to (34), $\langle s_1 \rangle$ and $\langle s_2 \rangle$ are expressed as
$$\langle s_1 \rangle_n = \begin{cases} \langle s_1 \rangle_0, & n = 0 \\ \dfrac{1}{\pi n} e^{jn\angle \langle s_1 \rangle_1} \sin \pi n \langle s_1 \rangle_0, & n \neq 0 \end{cases} \quad (80)$$

and
$$\langle s_2 \rangle_n = \begin{cases} \langle s_2 \rangle_0, & n = 0 \\ \dfrac{1}{\pi n} e^{jn\angle \langle s_2 \rangle_1} \sin \pi n \langle s_2 \rangle_0, & n \neq 0 \end{cases} \quad (81)$$

In the steady state, $\angle \langle s_1 \rangle_1$ is 0 if taking $s_1$ as the phase reference and $\angle \langle s_2 \rangle_1$ is $-\pi \langle s_1 \rangle_0 - \pi \langle s_2 \rangle_0$ because $t_{r2}$ equals $t_{f1}$. In CCM, $\langle s_2 \rangle_0$ is $1 - \langle s_1 \rangle_0$, and $\langle s_1 \rangle_0$ can be determined by iterating $v_c(t_{r1}) - k_v v_o(t_{r1})$ towards 0. In DCM, $\langle s_2 \rangle_0$ can be determined by iterating $i_L(t_{f2})$ towards 0, nesting within the iteration of $\langle s_1 \rangle_0$.

Given $\langle v_i \rangle$, $\langle s_1 \rangle$ and $\langle s_2 \rangle$, the steady-state values of $\langle i_L \rangle$, $\langle u_C \rangle$, and $\langle v_o \rangle$ can be solved from the linear equilibrium equation:
$$\begin{cases} \begin{bmatrix} [\![s_1]\!]\langle v_i \rangle - ([\![s_1]\!] + [\![s_2]\!])\langle v_o \rangle - jN\omega L \langle i_L \rangle \\ \langle i_L \rangle - \dfrac{1}{R}\langle v_o \rangle - jN\omega C \langle u_C \rangle \end{bmatrix} = 0 \\ \langle v_o \rangle = \dfrac{Rr}{R+r}\langle i_L \rangle + \dfrac{R}{R+r}\langle u_C \rangle \end{cases} \quad (82)$$

and $v_c(t_{r1}) - k_v v_o(t_{r1})$ and $i_L(t_{f2})$ can be calculated using $\langle v_c \rangle$, $\langle v_o \rangle$ and $\langle i_L \rangle$ according to (3) and (38) in the iterations.

### D. Experiment

The buck converter evaluation board used for the experiments is shown in Fig. 9. The parameters are shown in Table 4. The control voltage was generated by a signal generator and fed to the soft-start pin of the LM34919. In the experiment, the on-time was measured to be a constant 554 ns.

When the load resistance was 10 Ω, the circuit worked in CCM, as shown in Fig. 10 (a). The measured switching frequency was 813 kHz and the duty cycle was 45%, both higher than the theoretical values due to circuit losses. To compensate for the effect of losses, the input voltage used for modeling was reduced to 11.2 V. Fig. 10 (c) shows the waveforms derived from the corresponding steady-state solutions for the −49th to 49th moving Fourier coefficients. The derived waveforms, switching frequency, and duty cycle were consistent with the experimental results.

To obtain the frequency-domain characteristics, the control voltage was modulated with a modulation depth of 10 mV and a modulation frequency ranging from 100 kHz to 1.4 MHz, and the responses of the converter were recorded. For example, Fig. 10 (e) shows the responses captured at a modulation frequency of 800 kHz. From the spectrum of the response data, the value of the control voltage-to-output voltage transfer function at each modulation frequency was obtained, as shown in Fig. 10 (g). For comparison, Fig. 10 (g) also shows the transfer function curves predicted by the generalized average model built in this section and the describing function model [15]. The magnitude-frequency curves given by the two models were almost the same, but there was a significant difference near the switching frequency. The generalized average model predicted a peak but the describing function model predicted a valley. The experimental data enclosed by the dashed circle supported the prediction of the generalized average model. The difference in the phase-frequency curves given by the two models was mainly due to the fact that the describing function model did not consider the logic and drive delay, which was identified in the DCM experiment, as shown in Fig. 10 (b).

During the non-conduction period in DCM, the choke inductor and the parasitic capacitance of the switching devices formed a resonant tank, causing the inductor current and the output voltage to oscillate. At a certain oscillation valley, the detected output voltage was lower than the control voltage, triggering the next conduction period. In the experiment, it was captured that the rising time of the switch lagged behind the last oscillation valley by about 170 ns. Since the switching frequency was affected by the oscillation, the circuit could not enter a steady state in DCM. The steady-state solution and frequency characteristics in DCM were given by model calculation and circuit simulation rather than experiments, as shown in Fig. 10 (d) and (f). Fig. 10 (h) verifies the control voltage-to-output voltage transfer function predicted by the generalized average model built for DCM using the simulation data.

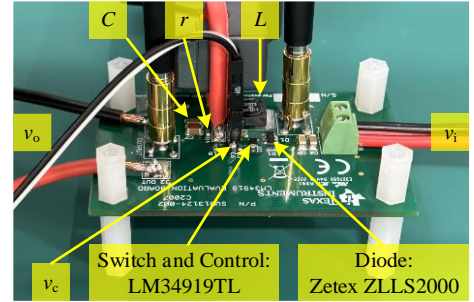

Fig. 9. Buck converter evaluation board (LM34919EVAL from TI).

Table 4. Buck converter parameters

| Symbol | Quantity | Value | |
|---|---|---|---|
| | | CCM | DCM |
| $L$ | Choke inductance | 15 μH | 15 μH |
| $C$ | Filter capacitance | 20 μF | 20 μF |
| $r$ | Series resistance | 0.39 Ω | 0.39 Ω |
| $R$ | Load resistance | 10 Ω | 100 Ω |
| $v_i$ | Input voltage | 12 V | 12 V |
| $v_c$ | Steady-state control voltage | 2.5 V | 2.5 V |
| $T_{on}$ | On-time | 554 ns | 554 ns |
| $T_d$ | Logic and drive delay | 170 ns | 170 ns |
| $f_s$ | Steady-state switching frequency | 813 kHz | *371 kHz |

*Simulation result

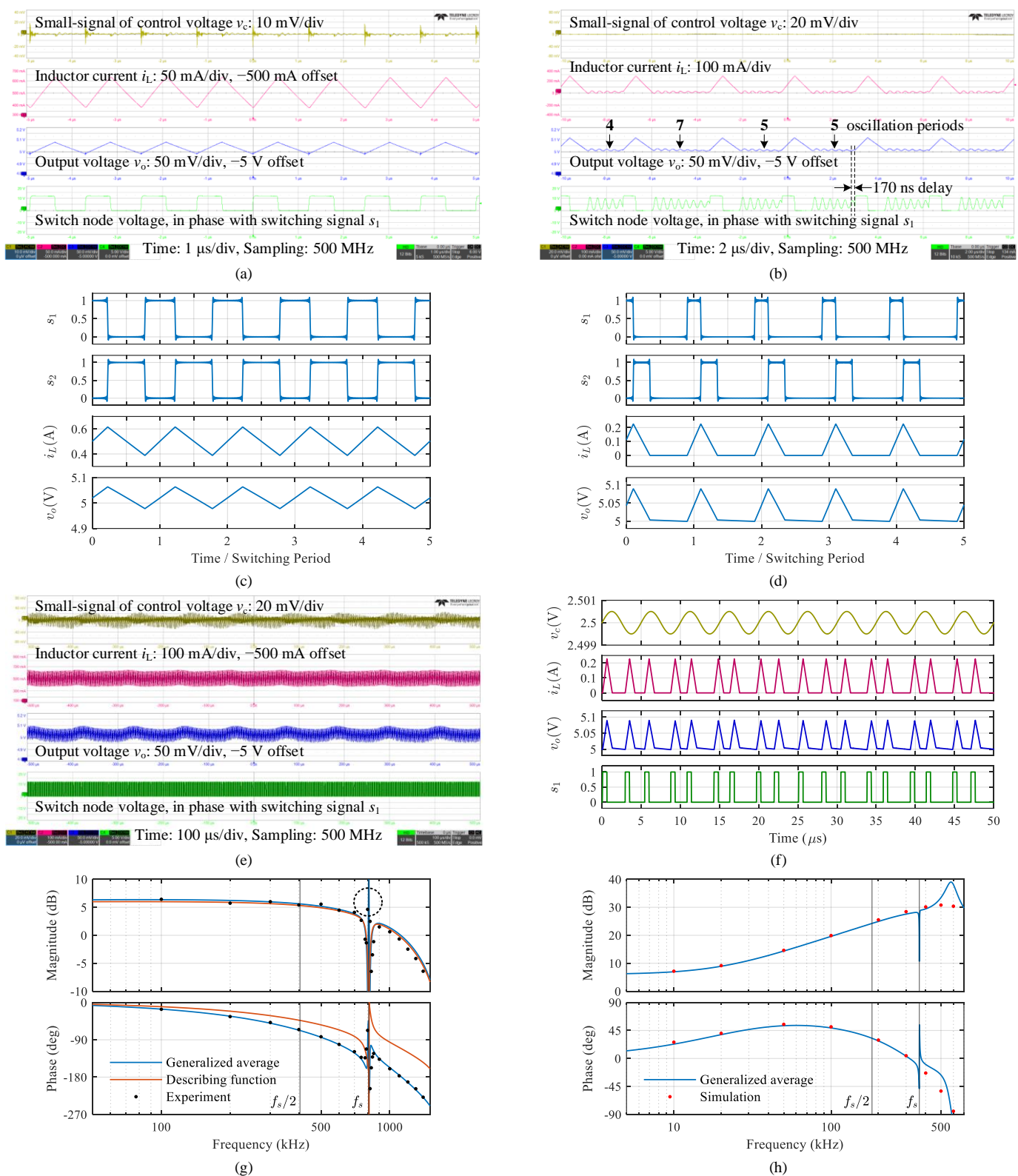

Fig. 10. Experimental, simulation, and calculation results of the buck converter: (a) steady-state waveforms captured in CCM, (b) waveforms captured in DCM, waveforms derived from the steady-state solutions for the moving Fourier coefficients in (c) CCM and (d) DCM, (e) responses captured when the control voltage was modulated with a modulation depth of 10 mV and a modulation frequency of 800 kHz in CCM, (f) simulation waveforms when the control voltage was modulated with a modulation depth of 0.5 mV and a modulation frequency of 200 kHz in DCM, control voltage-to-output voltage transfer function in (g) CCM and (h) DCM.

## VII. PFM Controlled LLC Converter

The LLC converter, as shown in Fig. 11, is a representative resonant converter widely used in isolated DC-DC conversion. It uses a three-element resonant tank for primary-side zero-voltage switching and secondary-side zero-current switching. The magnetizing inductance and leakage inductance of the transformer can provide two of the three resonant elements to achieve high power density [33]. LLC converters are typically controlled by the switching frequency using the bandpass characteristics of the resonant tank.

LLC converters are often modeled using the EDF method or the generalized averaging method that considers only the DC and fundamental harmonics [34, 35]. These models are not accurate when the waveform deviates significantly from sinusoidal. This section builds the generalized average model of LLC converter that is accurate from DC to above half the switching frequency.

### A. Generalized average model of the circuit

The state-space model of the circuit is

$$\begin{bmatrix} L_r & 0 & 0 & 0 \\ 0 & C_r & 0 & 0 \\ 0 & 0 & L_m & 0 \\ 0 & 0 & 0 & C_f \end{bmatrix} \frac{d}{dt} \begin{bmatrix} i_r \\ u_r \\ i_m \\ v_o \end{bmatrix}$$
$$= \begin{bmatrix} [1 - k^2(1 - s_3 - s_4)](s_1 v_i - u_r) - r(s_3 - s_4)v_o \\ i_r \\ k^2(1 - s_3 - s_4)(s_1 v_i - u_r) + r(s_3 - s_4)v_o \\ r(s_3 - s_4)(i_r - i_m) - \frac{1}{R}v_o \end{bmatrix} \quad (83)$$

where $L_r$, $C_r$, $L_m$, $C_f$, $R_L$, and $r$ are the resonant inductance, resonant capacitance, magnetizing inductance, filter capacitance, load resistance, and transformer turn ratio, respectively, $i_r$, $u_r$, $i_m$, $v_i$, and $v_o$ are the resonant current, resonant voltage, magnetizing current, input voltage, and output voltage, respectively, $k^2$ is given by

$$k^2 = \frac{L_m}{L_r + L_m} \quad (84)$$

(If $L_r$ is provided by the transformer leakage inductance, then $k$ is the coupling coefficient of the transformer.) $s_1$, $s_3$, and $s_4$ are the switching signals of primary-side switch $S_1$ and secondary-side diodes $D_3$ and $D_4$, respectively.

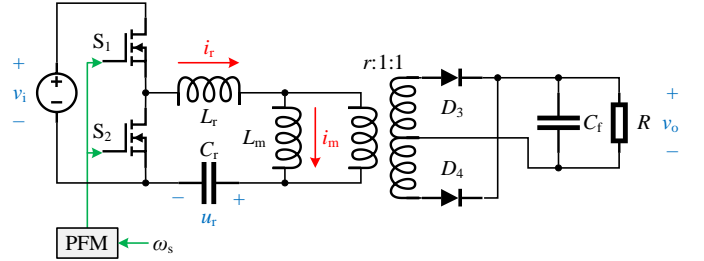

Fig. 11. PFM controlled LLC converter.

According to (27), the large-signal generalized average model of the circuit is derived from (83) and expressed as

*Equation at the bottom of this page* (85)

According to (31), the small-signal generalized average model of the circuit is derived from (85) and expressed as

*Equation at the bottom of this page* (86)

### B. Generalized average model of the switching signals

For PFM control, the rising and falling instants of $s_1$, i.e. $t_{r1}$ and $t_{f1}$, are the zero points of a carrier $c$ with 50% positive period. The phase angle $\alpha$ of $c$ depends on the time-varying switching frequency $\omega_s$ as

$$\alpha = \int (\omega_s - \omega) dt \quad (87)$$

Taking $\alpha$ as the input, the PFM translates to PSM, and the small-signal generalized average model of $s_1$ is the combination of

$$\widetilde{\langle s_1 \rangle} = \begin{bmatrix} \frac{\partial \langle s_1 \rangle}{\partial t_{r1}} & \frac{\partial \langle s_1 \rangle}{\partial t_{f1}} \end{bmatrix} \begin{bmatrix} \widetilde{t_{r1}} \\ \widetilde{t_{f1}} \end{bmatrix} \quad (88)$$

and

$$\begin{bmatrix} \widetilde{t_{r1}} \\ \widetilde{t_{f1}} \end{bmatrix} = \begin{bmatrix} \frac{\partial t_{r1}}{\partial \langle \alpha \rangle} \\ \frac{\partial t_{f1}}{\partial \langle \alpha \rangle} \end{bmatrix} \widetilde{\langle \alpha \rangle} \quad (89)$$

According to (35), for any $n \in \mathbb{Z}$,

$$\begin{bmatrix} L_r I & 0 & 0 & 0 \\ 0 & C_r I & 0 & 0 \\ 0 & 0 & L_m I & 0 \\ 0 & 0 & 0 & C_f I \end{bmatrix} \frac{d}{d\tau} \begin{bmatrix} \langle i_r \rangle \\ \langle u_r \rangle \\ \langle i_m \rangle \\ \langle v_o \rangle \end{bmatrix} = \begin{bmatrix} [\langle 1 \rangle - k^2(\langle 1 \rangle - \langle s_3 \rangle - \langle s_4 \rangle)] * (\langle s_1 \rangle * \langle v_i \rangle - \langle u_r \rangle) - r(\langle s_3 \rangle - \langle s_4 \rangle) * \langle v_o \rangle - jN\omega L_r \langle i_r \rangle \\ \langle i_r \rangle - jN\omega C_r \langle u_r \rangle \\ k^2(\langle 1 \rangle - \langle s_3 \rangle - \langle s_4 \rangle) * (\langle s_1 \rangle * \langle v_i \rangle - \langle u_r \rangle) + r(\langle s_3 \rangle - \langle s_4 \rangle) * \langle v_o \rangle - jN\omega L_m \langle i_m \rangle \\ r(\langle s_3 \rangle - \langle s_4 \rangle) * (\langle i_r \rangle - \langle i_m \rangle) - \frac{1}{R}\langle v_o \rangle - jN\omega C_f \langle v_o \rangle \end{bmatrix} \quad (85)$$

$$\begin{bmatrix} L_r I & 0 & 0 & 0 \\ 0 & C_r I & 0 & 0 \\ 0 & 0 & L_m I & 0 \\ 0 & 0 & 0 & C_f I \end{bmatrix} \frac{d}{d\tau} \begin{bmatrix} \widetilde{\langle i_r \rangle} \\ \widetilde{\langle u_r \rangle} \\ \widetilde{\langle i_m \rangle} \\ \widetilde{\langle v_o \rangle} \end{bmatrix} = \begin{bmatrix} -jN\omega L_r & -I + k^2(I - [\![s_3]\!] - [\![s_4]\!]) & 0 & -r([\![s_3]\!] - [\![s_4]\!]) \\ I & -jN\omega C_r & 0 & 0 \\ 0 & -k^2(I - [\![s_3]\!] - [\![s_4]\!]) & -jN\omega L_m & r([\![s_3]\!] - [\![s_4]\!]) \\ r([\![s_3]\!] - [\![s_4]\!]) & 0 & -r([\![s_3]\!] - [\![s_4]\!]) & -\frac{1}{R}I - jN\omega C_f \end{bmatrix} \begin{bmatrix} \widetilde{\langle i_r \rangle} \\ \widetilde{\langle u_r \rangle} \\ \widetilde{\langle i_m \rangle} \\ \widetilde{\langle v_o \rangle} \end{bmatrix}$$
$$+ \begin{bmatrix} [I - k^2(I - [\![s_3]\!] - [\![s_4]\!])][\![v_i]\!] & k^2([\![s_1]\!][\![v_i]\!] - [\![u_r]\!]) - r[\![v_o]\!] & k^2([\![s_1]\!][\![v_i]\!] - [\![u_r]\!]) + r[\![v_o]\!] \\ 0 & 0 & 0 \\ k^2(I - [\![s_3]\!] - [\![s_4]\!])[\![v_i]\!] & -k^2([\![s_1]\!][\![v_i]\!] - [\![u_r]\!]) + r[\![v_o]\!] & -k^2([\![s_1]\!][\![v_i]\!] - [\![u_r]\!]) - r[\![v_o]\!] \\ 0 & r([\![i_r]\!] - [\![i_m]\!]) & -r([\![i_r]\!] - [\![i_m]\!]) \end{bmatrix} \begin{bmatrix} \widetilde{\langle s_1 \rangle} \\ \widetilde{\langle s_3 \rangle} \\ \widetilde{\langle s_4 \rangle} \end{bmatrix} \quad (86)$$

$$\begin{cases} \dfrac{\partial \langle s_1 \rangle_n}{\partial t_{r1}} = -\dfrac{1}{T} e^{-jn\omega t_{r1}} \\ \dfrac{\partial \langle s_1 \rangle_n}{\partial t_{f1}} = \dfrac{1}{T} e^{-jn\omega t_{f1}} \end{cases} \quad (90)$$

According to (50), for any $m \in \mathbb{Z}$,

$$\begin{cases} \dfrac{\partial t_{r1}}{\partial \langle \alpha \rangle_m} = \dfrac{-e^{jm\omega t_{r1}}}{\omega} \\ \dfrac{\partial t_{f1}}{\partial \langle \alpha \rangle_m} = \dfrac{-e^{jm\omega t_{f1}}}{\omega} \end{cases} \quad (91)$$

On the secondary side, the rising and falling instants of $s_3$ and $s_4$, i.e. $t_{r3}, t_{f3}, t_{r4}$, and $t_{f4}$, are the zero points of $i_r - i_m$, and therefore, the small-signal generalized average model of $s_3$ and $s_4$ is the combination of

$$\begin{cases} \langle \widetilde{s_3} \rangle = \begin{bmatrix} \dfrac{\partial \langle s_3 \rangle}{\partial t_{r3}} & \dfrac{\partial \langle s_3 \rangle}{\partial t_{f3}} \end{bmatrix} \begin{bmatrix} \widetilde{t_{r3}} \\ \widetilde{t_{f3}} \end{bmatrix} \\ \langle \widetilde{s_4} \rangle = \begin{bmatrix} \dfrac{\partial \langle s_4 \rangle}{\partial t_{r4}} & \dfrac{\partial \langle s_4 \rangle}{\partial t_{f4}} \end{bmatrix} \begin{bmatrix} \widetilde{t_{r4}} \\ \widetilde{t_{f4}} \end{bmatrix} \end{cases} \quad (92)$$

and

$$\begin{cases} \begin{bmatrix} \widetilde{t_{r3}} \\ \widetilde{t_{f3}} \end{bmatrix} = \begin{bmatrix} \dfrac{\partial t_{r3}}{\partial \langle i_r \rangle} & \dfrac{\partial t_{r3}}{\partial \langle i_m \rangle} \\ \dfrac{\partial t_{f3}}{\partial \langle i_r \rangle} & \dfrac{\partial t_{f3}}{\partial \langle i_m \rangle} \end{bmatrix} \begin{bmatrix} \langle \widetilde{i_r} \rangle \\ \langle \widetilde{i_m} \rangle \end{bmatrix} \\ \begin{bmatrix} \widetilde{t_{r4}} \\ \widetilde{t_{f4}} \end{bmatrix} = \begin{bmatrix} \dfrac{\partial t_{r4}}{\partial \langle i_r \rangle} & \dfrac{\partial t_{r4}}{\partial \langle i_m \rangle} \\ \dfrac{\partial t_{f4}}{\partial \langle i_r \rangle} & \dfrac{\partial t_{f4}}{\partial \langle i_m \rangle} \end{bmatrix} \begin{bmatrix} \langle \widetilde{i_r} \rangle \\ \langle \widetilde{i_m} \rangle \end{bmatrix} \end{cases} \quad (93)$$

According to (35), for any $n \in \mathbb{Z}$,

$$\begin{cases} \dfrac{\partial \langle s_3 \rangle_n}{\partial t_{r3}} = -\dfrac{1}{T} e^{-jn\omega t_{r3}} \\ \dfrac{\partial \langle s_3 \rangle_n}{\partial t_{f3}} = \dfrac{1}{T} e^{-jn\omega t_{f3}} \\ \dfrac{\partial \langle s_4 \rangle_n}{\partial t_{r4}} = -\dfrac{1}{T} e^{-jn\omega t_{r4}} \\ \dfrac{\partial \langle s_4 \rangle_n}{\partial t_{f4}} = \dfrac{1}{T} e^{-jn\omega t_{f4}} \end{cases} \quad (94)$$

According to (41) and (48), for any $m \in \mathbb{Z}$,

$$\begin{cases} \dfrac{\partial t_{r3}}{\partial \langle i_r \rangle_m} = -\dfrac{\partial t_{r3}}{\partial \langle i_m \rangle_m} = \dfrac{-e^{jm\omega t_{r3}}}{\sum_{n=-\infty}^{+\infty} jn\omega \langle i_r - i_m \rangle_n e^{jn\omega t_{r3}}} \\ \dfrac{\partial t_{f3}}{\partial \langle i_r \rangle_m} = -\dfrac{\partial t_{f3}}{\partial \langle i_m \rangle_m} = \dfrac{-e^{jm\omega t_{f3}}}{\sum_{n=-\infty}^{+\infty} jn\omega \langle i_r - i_m \rangle_n e^{jn\omega t_{f3}}} \\ \dfrac{\partial t_{r4}}{\partial \langle i_r \rangle_m} = -\dfrac{\partial t_{r4}}{\partial \langle i_m \rangle_m} = \dfrac{-e^{jm\omega t_{r4}}}{\sum_{n=-\infty}^{+\infty} jn\omega \langle i_r - i_m \rangle_n e^{jn\omega t_{r4}}} \\ \dfrac{\partial t_{f4}}{\partial \langle i_r \rangle_m} = -\dfrac{\partial t_{f4}}{\partial \langle i_m \rangle_m} = \dfrac{-e^{jm\omega t_{f4}}}{\sum_{n=-\infty}^{+\infty} jn\omega \langle i_r - i_m \rangle_n e^{jn\omega t_{f4}}} \end{cases} \quad (95)$$

### C. Steady-state operating point

According to (34), $\langle s_1 \rangle$ and $\langle s_3 \rangle$ are expressed as

$$\langle s_1 \rangle_n = \begin{cases} \langle s_1 \rangle_0, & n = 0 \\ \dfrac{1}{\pi n} e^{jn \angle \langle s_1 \rangle_1} \sin \pi n \langle s_1 \rangle_0, & n \neq 0 \end{cases} \quad (96)$$

and

$$\langle s_3 \rangle_n = \begin{cases} \langle s_3 \rangle_0, & n = 0 \\ \dfrac{1}{\pi n} e^{jn \angle \langle s_3 \rangle_1} \sin \pi n \langle s_3 \rangle_0, & n \neq 0 \end{cases} \quad (97)$$

Due to the symmetrical conduction of the diodes, the phase difference between $s_3$ and $s_4$ is $\pi$, and therefore

$$\langle s_4 \rangle_n = \langle s_3 \rangle_n e^{jn\pi} \quad (98)$$

In the steady state, $\langle s_1 \rangle_0$ is 0.5 because the duty cycle of $s_1$ is 50%, $\angle \langle s_1 \rangle_1$ is 0 if taking $s_1$ as the phase reference. If the steady-state switching frequency $\omega$ is below the resonant frequency $\omega_r$ of $L_r$ and $C_r$, then both $\langle s_3 \rangle_0$ and $\angle \langle s_3 \rangle_1$ can be determined by iterating $i_r(t_{f3}) - i_m(t_{f3})$ towards 0 with the constraints that $t_{r3}$ equals $t_{r1}$ and $t_{r4}$ equals $t_{f1}$. If $\omega$ is above $\omega_r$, then $\langle s_3 \rangle_0$ is 0.5 and $\angle \langle s_3 \rangle_1$ can be determined by iterating $i_r(t_{f3}) - i_m(t_{f3})$ towards 0.

Given $\langle v_i \rangle$, $\langle s_1 \rangle$, $\langle s_3 \rangle$, and $\langle s_4 \rangle$, the steady-state values of $\langle i_r \rangle$, $\langle u_r \rangle$, $\langle i_m \rangle$, and $\langle v_o \rangle$ can be solved from the linear equilibrium equation:

*Equation at the bottom of this page*  (99)

and $i_r(t_{f3}) - i_m(t_{f3})$ can be calculated using $\langle i_r \rangle$ and $\langle i_m \rangle$ according to (3) and (38) in the iteration.

### D. Experiment

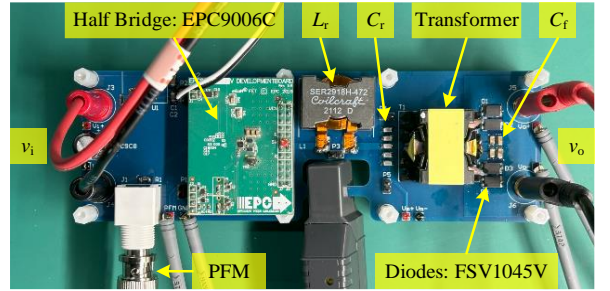

Fig. 12. LLC converter prototype.

Table 5. LLC converter parameters

| Symbol | Quantity | Value | |
|---|---|---|---|
| | | $f_s < f_r$ | $f_s > f_r$ |
| $L_r$ | Resonant inductance | 4.7 µH | 4.7 µH |
| $C_r$ | Resonant capacitance | 530 nF | 530 nF |
| $f_r$ | Resonant frequency | 101 kHz | 101 kHz |
| $L_m$ | Magnetizing inductance | 25 µH | 25 µH |
| $r$ | Transformer turn ratio | 2 | 2 |
| $C_f$ | Filter capacitance | 8 µF | 8 µF |
| $R_L$ | Load resistance | 2 Ω | 2 Ω |
| $v_i$ | Input voltage | 48 V | 48 V |
| $f_s$ | Steady-state switching frequency | 80 kHz | 120 kHz |

$$\begin{bmatrix} [\langle 1 \rangle - k^2(\langle 1 \rangle - \langle s_3 \rangle - \langle s_4 \rangle)] * (\langle s_1 \rangle * \langle v_i \rangle - \langle u_r \rangle) - r(\langle s_3 \rangle - \langle s_4 \rangle) * \langle v_o \rangle \\ \langle i_r \rangle \\ k^2(\langle 1 \rangle - \langle s_3 \rangle - \langle s_4 \rangle) * (\langle s_1 \rangle * \langle v_i \rangle - \langle u_r \rangle) + r(\langle s_3 \rangle - \langle s_4 \rangle) * \langle v_o \rangle \\ r(\langle s_3 \rangle - \langle s_4 \rangle) * (\langle i_r \rangle - \langle i_m \rangle) - \dfrac{1}{R} \langle v_o \rangle \end{bmatrix} = 0 \quad (99)$$

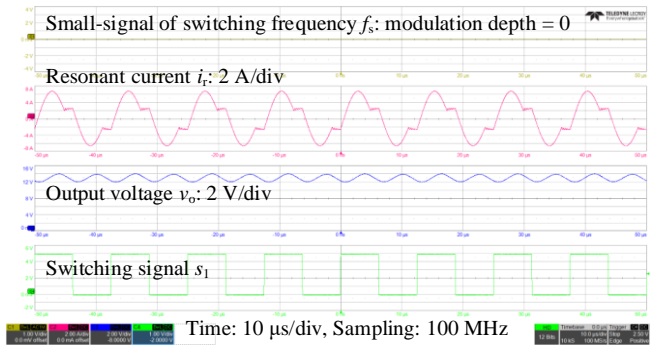
(a)

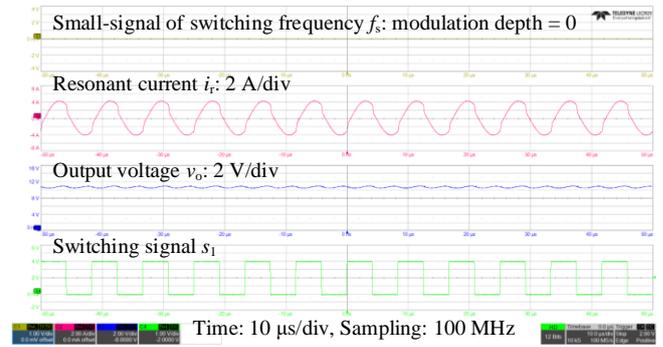
(b)

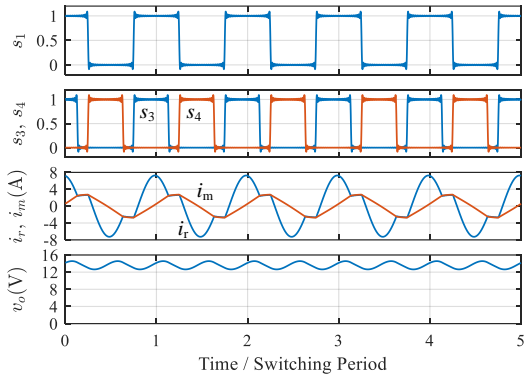
(c)

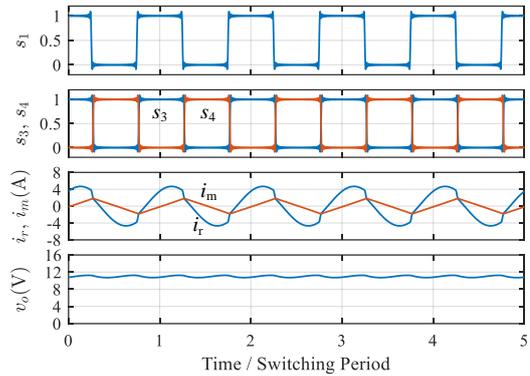
(d)

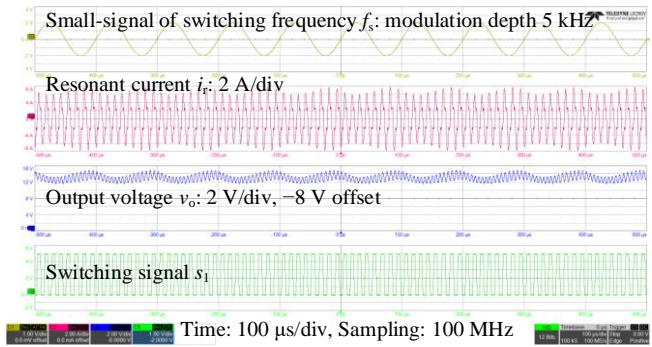
(e)

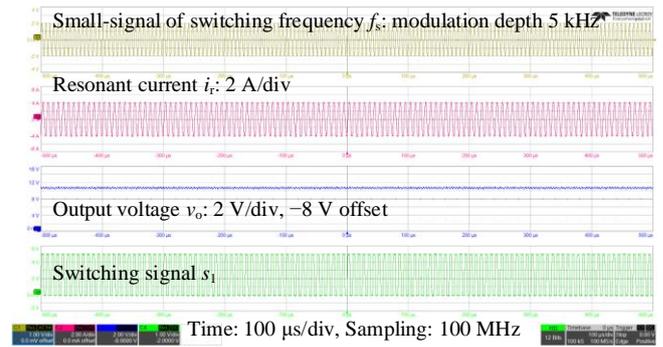
(f)

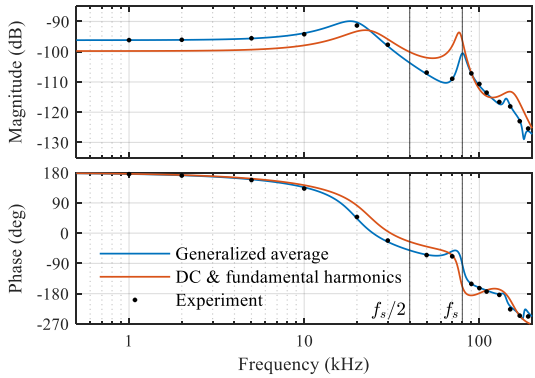
(g)

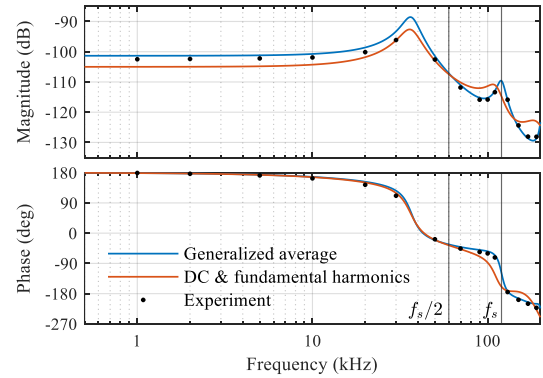
(h)

Fig. 13. Experimental and calculation results of the LLC converter: steady-state waveforms captured when the switching frequency was (a) 80 kHz and (b) 120 kHz, waveforms derived from the steady-state solutions for the moving Fourier coefficients when the switching frequency was (c) 80 kHz and (d) 120 kHz, responses captured when the switching frequency was modulated with a modulation depth of 5 kHz (e) around 80 kHz at a modulation frequency of 10 kHz and (f) around 120 kHz with a modulation frequency of 150 kHz, switching frequency-to-output voltage transfer function when the switching frequency was around (g) 80 kHz and (h) 120 kHz.

The LLC converter prototype used for the experiments is shown in Fig. 12. The parameters are shown in Table 5. The converter was controlled by a PFM signal generated by a signal generator. Experiments were performed at steady-state switching frequencies of 80 kHz (below the resonant frequency) and 120 kHz (above the resonant frequency) respectively.

Fig. 13 (a) and (b) show the steady-state waveforms captured at the two switching frequencies. Fig. 13 (c) and (d) show the waveforms derived from the corresponding steady-state solutions for the −49th to 49th moving Fourier coefficients, which were consistent with the captured waveforms.

To obtain the frequency-domain characteristics, the frequency of the PFM signal was modulated with a modulation depth of 5 kHz and a modulation frequency ranging from 1 kHz to 190 kHz, and the responses of the converter were recorded. For example, Fig. 13 (e) and (f) show the responses captured when the switching frequency was modulated around 80 kHz at a modulation frequency of 10 kHz and around 120 kHz at a modulation frequency of 150 kHz, respectively. From the spectrum of the response data, the value of the switching frequency-to-output voltage transfer function at each modulation frequency was obtained, as shown in Fig. 13 (g) and (h). For comparison, Fig. 13 (g) and (h) also show the transfer function curves predicted by the generalized average model built in this section and the widely used fundamental harmonic model [34, 35]. The generalized average model is more accurate than the fundamental harmonic model, especially when the modulation frequency is higher than half the switching frequency.

## VIII. Conclusion

This paper uses the properties of moving Fourier coefficients to develop the generalized averaging method, and enables the method to describe the dynamics of multiple harmonics in power electronic converters with concise mathematical formulas, allowing the effective range of the model to break through the limit of half the switching frequency. The paper also proposes generalized average models for various switching signals, including PWM, PSM, PFM, and state-dependent switching signals, so that circuits and modulators/controllers can be modeled separately and then combined flexibly. Using the Laplace transform of moving Fourier coefficients, the frequency-domain relationship between a signal and its moving Fourier coefficients (which include the moving average) is clarified, and the coupling of signals and their sidebands at different frequencies is clearly and conveniently described by a transfer function matrix in an LTI system framework.

The steps to build a generalized average model of a power electronic converter are summarized as follows.
1) Build the time-variant state-space model of the circuit with switching signals as parameters;
2) Derive the large-signal generalized average model of the circuit from the state-space model;
3) Calculate the steady-state operating point using the linear equilibrium equation and iterations if necessary;
4) Derive the small-signal generalized average model of the circuit at the steady-state operating point;
5) Derive the small-signal generalized average model of the switching signals according to the modulation/control scheme;
6) Combine the models of the circuit and the switching signals linearly;
7) Calculate the transfer function matrix between moving Fourier coefficients of signals of interest using the combined model.

The final output of the modeling is the small-signal model and the transfer function matrix, which describe not only the relationship between moving Fourier coefficients, but also the relationship between original signals and their sidebands. The intermediate output of the modeling is the large-signal model, which can accurately describe the transient response of the circuit from the zero state to the steady state, so that the steady-state operating point (and the steady-state waveforms with arbitrary accuracy) can be obtained through several times of linear calculation, which is much faster than conventional simulation methods.

The developed generalized average model is theoretically infinite-dimensional, but the executable calculations must contain a finite number of moving Fourier coefficients. The truncation of moving Fourier coefficients distorts the calculated steady-state waveforms and introduces errors in the derived transfer functions. The experimental results in this paper show that a very accurate model can be obtained by taking into account dozens of moving Fourier coefficients. The quantitative relationship between the truncation and the error is an issue that could be studied in the future.

The paper gives three representative modeling examples and shows experimental results in different operating modes, demonstrating that the developed generalized averaging method has broad applicability. Compared with existing models, the proposed models have higher accuracy, especially in the frequency range close to or above half the switching frequency. In addition to the examples, the method can be applied to a wide variety of power electronic converters, such as DC-DC converters with different control schemes, switched capacitor converters, wireless power transfer systems, dual-active bridge converters, and grid-connected converters.